\documentclass[fleqn]{cas-dc}
\pdfoutput=1
\usepackage[authoryear,longnamesfirst]{natbib}

\usepackage{lineno}
\usepackage{amsmath,amssymb}

\usepackage{journals} 
\usepackage{booktabs}
\usepackage{longtable}

\newcommand{\abar}{\tilde{\alpha}}
\newcommand{\tbar}{\tilde{T}}
\newcommand{\rbar}{\tilde{\rho}}
\newcommand{\gbar}{\tilde{g}}
\newcommand{\sbar}{\tilde{s}}

\newcommand{\hmag}{\mathcal{H}_m}
\newcommand{\hstrat}{\mathcal{H}_s}
\newcommand{\rins}{\mathcal{R}_i}
\newcommand{\routs}{\mathcal{R}_o}
\newcommand{\rcs}{\mathcal{R}_c}
\newcommand{\cyl}{\varsigma}

\DeclareMathOperator{\argmax}{\arg\!\max}

\def\vec#1{\ensuremath{\mathchoice{\mbox{\boldmath$\displaystyle#1$}}
{\mbox{\boldmath$\textstyle#1$}}
{\mbox{\boldmath$\scriptstyle#1$}}
{\mbox{\boldmath$\scriptscriptstyle#1$}}}}
\def\tens#1{\ensuremath{\mathsf{#1}}}

\begin{document}
\let\WriteBookmarks\relax
\def\floatpagepagefraction{1}
\def\textpagefraction{.001}


\shorttitle{Stable stratification promotes multiple zonal jets in a 
turbulent Jovian dynamo model}

\title[mode=title]{Stable stratification promotes multiple zonal jets in a 
turbulent Jovian dynamo model}

\shortauthors{T.~Gastine and J.~Wicht}
\author[1]{T.~Gastine}[orcid=0000-0003-4438-7203]
\author[2]{J.~Wicht}[orcid=0000-0002-2440-5091]

\affiliation[1]{organization={Universit\'e de Paris, Institut de Physique du 
Globe de Paris},addressline={UMR 7154 CNRS, 1 rue 
Jussieu},city={F-75005, Paris},country=France}
\affiliation[2]{organization={Max Planck Institut f\"ur 
Sonnensytemforschung},addressline={Justus-von-Liebig-Weg 3},city={37077, 
G\"ottingen},country=Germany}

\cormark[1]
\cortext[1]{Corresponding author. E-mail: 
\href{mailto:gastine@ipgp.fr}{gastine@ipgp.fr}}


\begin{abstract}
The ongoing NASA's Juno mission puts new constraints on the internal dynamics 
of Jupiter. Data gathered by its onboard magnetometer reveal a  
dipole-dominated surface magnetic field accompanied by strong localised 
magnetic flux patches. The gravity measurements indicate 
that the fierce surface zonal jets extend several thousands of kilometers 
below the cloud level before rapidly decaying below $0.94-0.96\,R_J$, $R_J$ 
being the mean Jovian radius at the one bar level. Jupiter's 
internal structure can be grossly decomposed in two parts: (\textit{i}) an outer 
layer filled with a mixture of molecular hydrogen and helium where the zonal 
flows are thought to be driven; (\textit{ii})
an inner region where hydrogen becomes metallic and dynamo action is
expected to sustain the magnetic field. Several internal models
however suggest a more intricate structure with a thin intermediate region in 
which helium would segregate from hydrogen, forming a 
compositionally-stratified layer. Here, we develop the first global Jovian 
dynamo which incorporates an intermediate stably-stratified layer between 
$0.82\,R_J$ and $0.86\,R_J$.
Using much lower diffusivities than previous models enables us to more clearly 
separate the dynamics of the metallic core and the molecular envelope.
Analysing the energy balance reveals that the magnetic energy is almost one 
order of magnitude larger than kinetic energy in the metallic region, while 
most of the kinetic energy is pumped into zonal motions in the molecular 
envelope. Those result from 
the different underlying force hierarchy with a triple balance between 
Lorentz, Archimedean and ageostrophic Coriolis forces in the metallic core and
inertia, buoyancy and ageostrophic Coriolis forces controlling the external 
layers.
The simulation presented here is the first to demonstrate that multiple zonal 
jets and dipole-dominated dynamo action can be consolidated in a global 
simulation. 
The inclusion of a stable layer is a necessary ingredient that allows zonal 
jets to develop in the outer envelope without contributing to the dynamo action 
in the deeper metallic region. 
Stable stratification however also smooths out the small-scale features of the 
magnetic field by skin effect, yielding a too-dipolar surface field as compared 
to the observations. These constraints suggest that possible stable layers in 
Jupiter should be located much closer to the surface ($0.9-0.95\,R_J$).
\end{abstract}


\maketitle

\section{Introduction}

The banded structures observed at Jupiter's surface correlate with strong 
prograde (or eastward) and retrograde (or westward) winds. A strong prograde 
equatorial jet reaching $150$~m/s extends over $\pm 15^\circ$ latitude. 
It is flanked by alternating jets with weaker amplitudes around 
$10-20$~m/s up to the polar regions. 
The depth to which those winds penetrate into Jupiter 
has been debated intensely over the last 
decades \citep[for a review, see][]{Vasavada05}. In the ``weather layer'' 
scenario, the zonal jets are confined to a thin layer close to the cloud levels 
\citep[e.g.][]{Cho96,Lian10}, while under the ``deep convection'' hypothesis 
the zonal winds could penetrate deep over $10^3$ to $10^4$~km 
\citep[e.g.][]{Busse76,Christensen02,Heimpel05,Jones09}. Those two end-member 
scenarios also differ in the nature of the physical mechanism responsible for 
sustaining the jets. Possible candidates range from shallow moist convection at 
the cloud level to deep convective motions in Jupiter's interior.
For both physical forcings, rapid rotation is instrumental
for providing a statistical correlation that allows feeding energy from small 
scale convection in the larger scale jets \citep{Rhines75}.
Because of rapid rotation and the associated Taylor-Proudman theorem, the 
jets could penetrate deep into the molecular envelope even when they are only 
driven in a shallow weather layer \citep{Showman06}.

Determining the actual depth of the Jovian zonal jets is one of the 
main goals of the ongoing NASA Juno mission \citep{Bolton17}.
Using Juno's gravity measurements \citep{Iess18}, \cite{Kaspi18} infer that the 
equatorially-antisymmetric component of the zonal jets are reduced to an 
amplitude of $1\%$ of their surface values $3000$~km below the one bar level. 
However, the interpretation of gravity perturbations in terms of zonal flows is 
complicated and other zonal flow profiles could be envisioned 
\citep[e.g.][]{Kong18,Wicht20a,Galanti20}.

Several additional arguments favour comparable 
depths of $3000$ to $4000$~km. A first indication comes from studies of 
rapidly-rotating convection in thin spherical shells. Such numerical models 
have been developed to focus on the dynamics of the molecular envelope 
of the gas giants. They succeed in reproducing several 
key features of the observed zonal flow pattern such as a dominant prograde 
equatorial jet \citep[e.g.][]{Christensen02}, multiple jets of alternated 
directions \citep{Heimpel05,Jones09,Gastine14}, or the formation 
of large scale vorticies \citep{Heimpel16}. The width of the main prograde 
equatorial jet directly depends on the thickness of the simulated 
spherical shell \citep[e.g.][]{Heimpel07}. Best agreement with Jupiter is 
obtained when the lower boundary is set to $0.95\,R_J$.

\begin{figure}
 \centering
 \includegraphics[width=8.3cm]{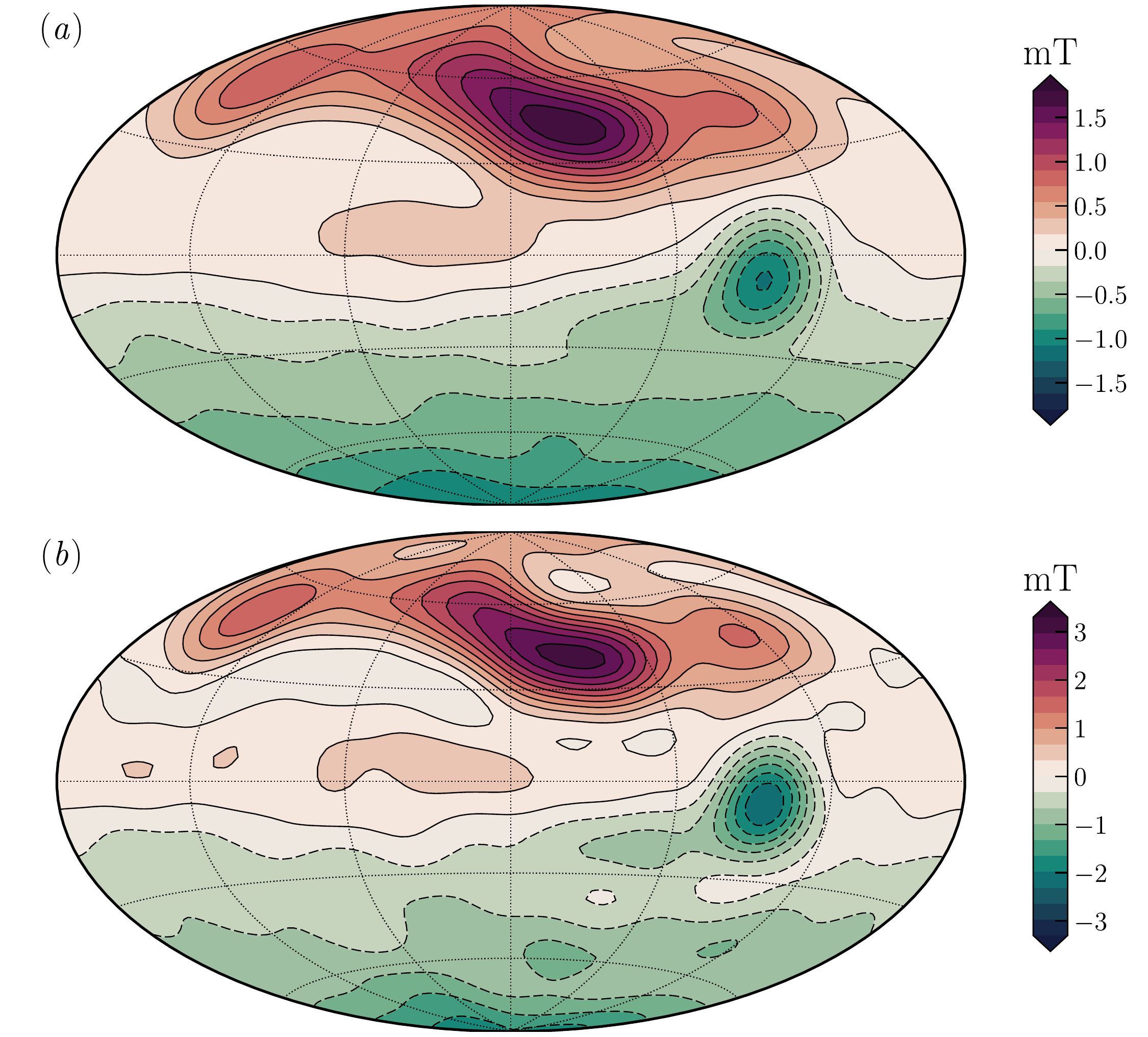}
 \caption{ Hammer projections of the
  radial component of the magnetic field at the surface of Jupiter (upper 
panel) and at $0.9\,R_J$ (lower panel). These maps have been recontructed using 
the JRM09 Jovian field model by \cite{Connerney18}.}
 \label{fig:BrJup}
\end{figure}

Another set of constraints on the zonal winds depth comes from the 
Jovian magnetic field. Using Juno's first nine orbits, \cite{Connerney18} have 
constructed the JRM09 internal field model up to the harmonic degree $\ell=10$ 
shown in Fig.~\ref{fig:BrJup}. The surface field (upper panel) is dominated by 
a tilted dipole and features intense localised flux concentrations. 
The downward continuation of the surface field to $0.9\,R_J$ 
(Fig.~\ref{fig:BrJup}\textit{b}) reveals an 
intricate field morphology with 
clear differences between the northern and southern hemispheres. In the 
northern hemisphere, the field is strongly concentrated in a latitudinal band, 
while the southern hemisphere is dominated by a pronounced field concentration 
just below the equator \citep{Moore18}.

A comparison of Juno's measurements with magnetic data from previous 
space missions, such as Pioneer or Voyager, shows only mild changes over a time 
span of 45 years \citep{Ridley16,Moore19}. This suggests an upper bound for the 
jet speed of roughly $1$~cm/s at a depth where magnetic effects start to 
matter at about $0.94\,R_J$.

Different lines of arguments therefore suggest a lower boundary for the jets 
located around $0.94-0.96\,R_J$. Which mechanism could possibly quench the jets 
in this depth range? Two alternatives have been suggested so far: Lorentz forces 
or a stably stratified layer.

Lorentz forces rely on electric currents and thus depend on the electrical 
conductivity. Experimental data 
\citep[e.g.][]{Weir96,Nellis99,Knudson18} and \textit{ab initio} simulations 
\citep[see][and references therein]{French12,Knudson18} indicate that the 
electrical conductivity increases at at super-exponential rate with depth due to 
the ionization of molecular hydrogen. At pressures of about one Mbar, however, 
hydrogen assumes a metallic state and the conductivity increases much more 
mildly. Here we use a model based on the \textit{ab initio} simulations by 
\cite{French12}, which puts the transition to metallic hydrogen at about 
$0.9\,R_J$.
However, many aspects of the conductivity profile remain debated. This includes 
the depth of the phase transition and the question of whether it is a first 
order or a gradual second order transition \citep[for a review 
see][]{Stevenson20}. 

A key parameter for estimating dynamo action is the magnetic Reynolds 
number $Rm$ which quantifies the ratio of induction and magnetic diffusion. In 
the outer envelope where the electrical conductivity increases 
extremely steeply, \cite{Liu08} showed that $Rm = U_z d_\sigma \sigma 
\mu_0 $ provides a more appropriate definition of the magnetic Reynolds number 
associated with zonal motions \citep[see also][]{Cao17}. Here $U_z$ is the 
typical zonal flow velocity, $\mu_0$ the vacuum permeability and 
$d_\sigma=|\partial \ln \sigma / \partial r|^{-1}$ the electrical
conductivity scale height. As long as $Rm$ remains smaller than unity, the 
zonal flows merely modify the field that is produced in the deeper 
interior \citep{Wicht19a}. In Jupiter, this region extends 
down to about $0.96\,R_J$ \citep{Wicht19}.
Lorentz forces then simply scale with $\sigma$  \citep{Wicht19a} and thus 
remain negligible in the very outer region but kick in abruptly at a certain 
depth.

While this suggest that Lorentz forces are a good candidate for quenching the 
jets, several numerical simulations reveal a different picture.
Instead of producing multiple alternating jets as the non-magnetic models, 
global dynamo simulations that adopt Jupiter's electrical conductivity profile  
only feature one main prograde equatorial jet 
aligned with the rotation axis, that mostly resides in the outer weakly 
conducting region.
Strong azimuthal Lorentz forces in the metallic interior
suppress zonal motions along the axis of rotation
and kill or significantly brake all other jets
\citep{Heimpel11,Duarte13,Jones14,Gastine14a,Dietrich18,Duarte18}.
Instead of explaining the 
observed depth, Lorentz forces seem to yield an unrealistic jet amplitude 
and structure \citep{Christensen20}.

Another candidate that could prevent
jets from penetrating deeper is a stably stratified layer that would 
inhibit the convective mixing. 
The Juno gravity observations suggest that Jupiter consists of several distinct 
layers: (\textit{i}) an outer envelope with reduced 
He (and Ne) abundance compared to the primordial solar value, 
(\textit{ii}) an intermediate envelope with a higher He abundance and possibly a 
lower abundance of heavier elements, 
(\textit{iii}) a deeper interior sometimes called a diluted core with an 
increased heavier element abundance, and (\textit{iv}) possibly a denser core 
\citep{Wahl17,Debras19,Stevenson20}. 

Stable stratification could help to explain how the different layers formed 
and were preserved over time. In gas giant planets, such stable 
layers could possibly occur when helium segregates from hydrogen due to its 
poor miscibility \citep[e.g.][]{Stevenson80,Lorenzen11}.
Below a critical temperature, helium tends to separate from hydrogen and 
forms droplets that rain towards the interior. This leads to helium depletion of 
the outer envelope and leaves a helium stably-stratifying gradient that 
separates the outer envelope from the interior. However, it remains unclear 
whether this process has already started in Jupiter 
\citep{Militzer16,Schottler18}.
If so, estimates put the upper boundary of the related stable layer around 
$1$~Mbar, which rougly corresponds to $0.9\,R_J$. 
The recent interior models by \cite{Debras19} put the upper boundary of 
the stable layer at $0.93\,R_J$ and the lower bound somewhere between 
$0.8\,R_J$ and $0.9\,R_J$.

In the limit of rapid rotation, the dynamical influence of a stably-stratified 
layer (hereafter SSL) depends on the ratio of the 
Brunt-V\"ais\"al\"a frequency $N$ to the rotation rate $\Omega$. 
Using a linear model of non-magnetic rotating convection,
\cite{Takehiro01} have shown that the distance of penetration $\delta$ of a 
convective feature of size $d_c$ into a stratified layer follows $\delta \sim 
(N/\Omega)^{-1}\,d_c$. Numerical models by \cite{Gastine20} showed that 
this scaling still holds in nonlinear dynamo models.
The penetration of zonal flows into such layers is more intricate since it 
directly depends on the thermal structure at the edge of the SSL 
\citep[e.g.][]{Showman06}. The global numerical models of 
solar-type stars by \cite{Brun17} show that the zonal motions do not penetrate 
into the stably-stratified interior when $N /\Omega \gg 1$ \citep[see 
also][]{Browning04,Augustson16}.
This ratio, however, remains poorly known in Jupiter's interior. The internal 
models by \cite{Debras19} suggest $1 \leq N/\Omega \leq 3$.
\citep{Christensen20}.
Considering simplified 2-D axisymmetric numerical models where the zonal 
flows are forced by an analytical source term, \cite{Christensen20} claim
that stable stratification alone is not sufficient to brake the geostrophic 
zonal winds. 
They suggest that weak Lorentz forces drive a weak meridional flow 
that penetrate the upper edge of the SSL, encountering the strong 
stable stratification. This in turn alters the latitudinal entropy 
structure that explains the quenching of the jets according to the thermal 
wind balance.

Here we adopt the idea of a stably-stratified sandwich layer and, for the first 
time, study its impact on the zonal jets  and overall dynamics in a full 3-D
global dynamo simulation.
The paper is organised as follows. Numerical model and methods are detailed in 
\S~\ref{sec:models}. Section~\ref{sec:results} is dedicated to the description 
of the results, while the implications for Jupiter are further discussed in 
\S~\ref{sec:disc}.

\section{Model and methods}

\label{sec:models}

\subsection{Defining a non-adiabatic reference state}

We consider a magnetohydrodynamic simulation of a conducting fluid in a 
spherical shell of radius ratio $r_i/r_o$ rotating at a constant rotation rate 
$\Omega$ about the $z$-axis. We adopt the so-called ``Lantz-Braginsky-Roberts'' 
anelastic approximation of the Navier-Stokes equations introduced by 
\cite{Brag95} and \cite{Lantz99}. It allows the incorporation of the 
radial dependence 
of the background state while filtering out the fast acoustic waves that would 
otherwise significantly hamper the timestep size. Within the anelastic 
approximation, one actually solves for small perturbations around a 
background state that is frequently assumed to be well-mixed and adiabatic 
\citep[e.g.][]{Jones11,Verhoeven15}.

Here we follow a slightly different approach. Since we aim at modelling the 
effects of a stably stratified layer at the top of the metallic region, we 
define a reference state that can depart from the adiabat.
This is a common approach in solar convection models that
incorporate both the radiative core and the convective envelope 
\citep[e.g.][]{Alvan14}. Practically, this implies that any physical quantity 
$x$ is expanded in spherical coordinates ($r,\theta,\phi)$ as follows

\begin{equation}
 x(r,\theta,\phi,t)=\tilde{x}(r)+x'(r,\theta,\phi,t),
\end{equation}
where the tilde denotes the spherically-symmetric and static
background state, while the primes correspond to fluctuations about this mean.
To ensure the validity of the anelastic approximation 
when using a non-adiabatic reference state, the perturbations should remain 
small as compared to the background state 
\citep[e.g.][]{Gough69}, i.e.

\[
 \dfrac{| x'|}{| x|} \ll 1,\quad \forall (r,\theta,\phi,t)\,.
\]
In the following, we adopt a dimensionless formulation of the MHD equations. 
Starting with the background reference state, the physical quantities such as 
the background density $\rbar$, temperature $\tbar$, gravity $\gbar$ and entropy 
gradient $\mathrm{d}\sbar/\mathrm{d} r$ are non-dimensionalised with respect 
to their value at the outer radius $r_o$. We adopt the spherical shell gap 
$d=r_o-r_i$ as the reference lengthscale.

\begin{figure}
 \centering
 \includegraphics[width=8.3cm]{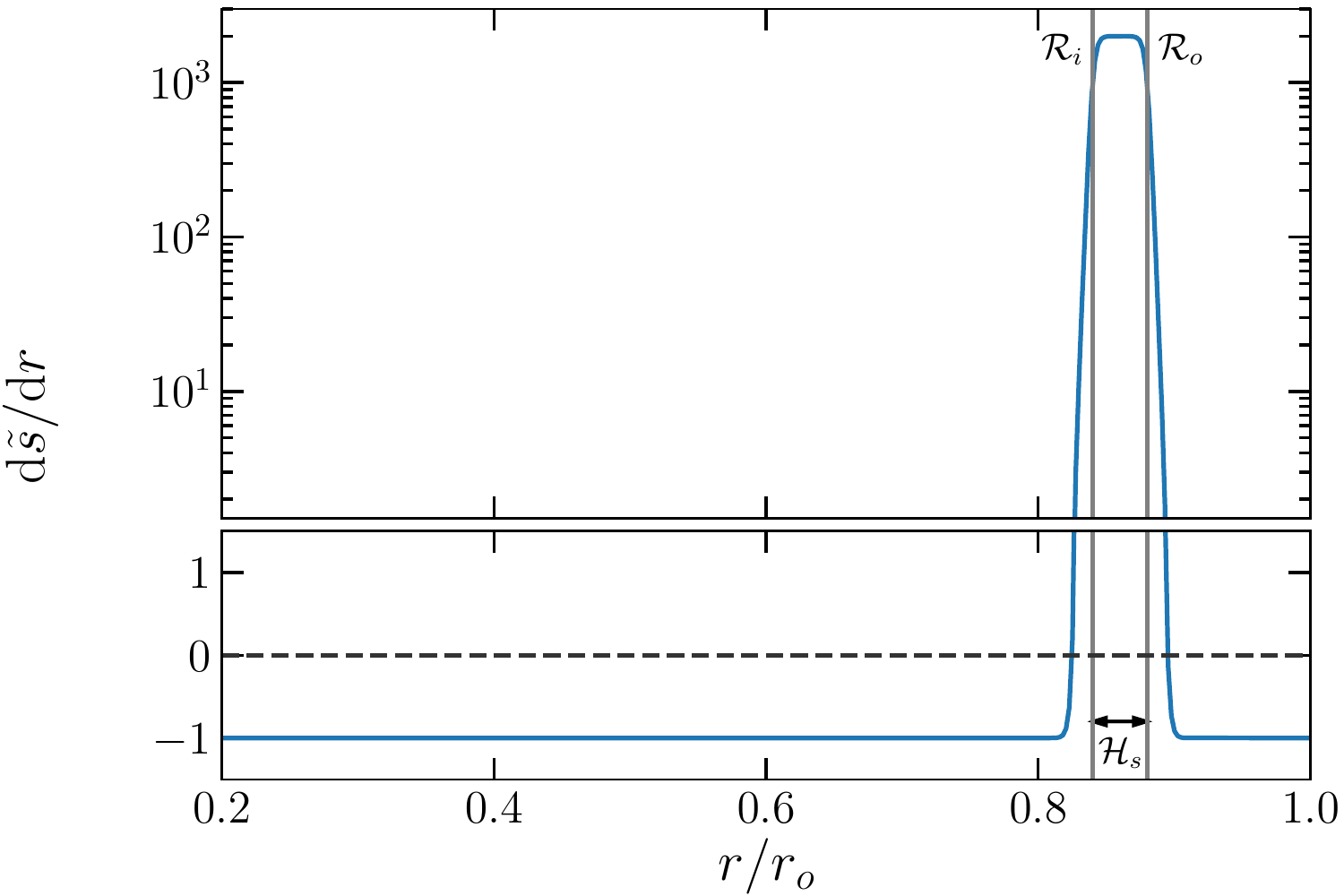}
 \caption{Background entropy gradient $\mathrm{d} 
\sbar/\mathrm{d} r$ as a function of the normalised radius $r/r_o$ as defined 
by Eq.~(\ref{eq:entropy}) with $\hstrat=0.05$, $\Gamma_s=2000$, 
$\rins/r_o=0.84$, $\routs/r_o=0.88$ and $\zeta_s=200$. The two vertical solid
lines mark the boundaries of the SSL $\rins$ and $\routs$. The horizontal 
dashed line corresponds to the neutral stratification $\mathrm{d} 
\sbar/\mathrm{d} r=0$, which delineates the separation between super adiabatic 
and stable stratification. To highlight the values of the profile in the 
convective regions, the $y$ axis has been split into logarithmic scale when 
$\mathrm{d}\sbar/\mathrm{d} r > 1.5$ (upper panel) and linear scale for 
the values between $-1.5$ and $1.5$ (lower panel).}
  \label{fig:dsdr}
\end{figure}

To precisely control the location and the degree of stratification of the SSL, a 
possible approach consists in prescribing the functional form of the background 
entropy gradient $\mathrm{d}\sbar/\mathrm{d}r$ \citep[see for instance ][for 
geodynamo models]{Takehiro01,Gastine20}. Regions with a 
negative gradient $\mathrm{d}\sbar/\mathrm{d} r < 0$ are super-adiabatic and 
hence prone to harbour convective motions, while the fluid layers 
with $\mathrm{d}\sbar/\mathrm{d} r > 0$ are stably 
stratified. In the following, we assume a constant degree of 
stratification 
$\mathrm{d}\sbar/\mathrm{d} r = \Gamma_s$ between the radii $\rins$ 
and $\routs$ and a constant dimensionless negative gradient 
$\mathrm{d}\sbar/\mathrm{d} r = -1$ in the surrounding convective layers. 
Those regions are then smoothly connected with $\tanh$ 
functions centered at $\rins$ and $\routs$:

\begin{equation}
\dfrac{\mathrm{d}\sbar}{\mathrm{d}r}=\dfrac{1+\Gamma_s}{4}\left[1+f_{\rins}
(r)\right ]\left[1-f_{\routs}(r)\right]-1,
\label{eq:entropy}
\end{equation}
where 
\[
f_{a}(r)=\tanh[\zeta_s(r-a)],\quad \routs=\rins+\hstrat,
\]
$\hstrat$ is the thickness of the stably stratified layer and $\zeta_s$ the 
stiffness of the transition. 
As we will see below, the degree of stratification $\Gamma_s$ can be directly 
related to the value of the Brunt-V\"ais\"al\"a frequency of the 
stably-stratified layer.
Figure~\ref{fig:dsdr} shows the radial profile  of $\mathrm{d}\sbar /\mathrm{d} 
r$ employed in this study. It features a stably-stratified layer between the 
radii 
$\rins=0.84\,r_o$ and $\routs=0.88\,r_o$, which correspond to $\hstrat=0.05$. 
The degree of stratification is set to $\Gamma_s=2000$, while the 
stiffness of the transition is $\zeta_s=200$. The 
location and thickness of the SSL have been chosen according to the internal 
models by \cite{Militzer16} and \cite{Wahl17}.
Because of the finite size of the transitions, $\mathrm{d}\sbar/\mathrm{d} r$ 
changes sign before (after) $\rins$ ($\routs$), yielding a stably-stratified 
layer with an effective thickness larger than $\hstrat$.
Imposing a background entropy gradient coming from stellar evolution 
models is commonly used in simulations of stellar interior dynamics 
\citep[e.g.][]{Browning04,Augustson16,Brun17} for introducing a 
stably-stratified region.
In absence of a more realistic entropy profile coming from internal models of 
Jupiter, we adopt here a parametrized background entropy gradient.
While being convenient, it lacks a proper physical justification and simply 
maintains the stratification by introducing an effective entropy or heat sink. 
A more realistic distribution of entropy or heat sources in the convective  
layer of Jupiter is discussed by \cite{Jones14}.

Once the background entropy gradient has been specified, 
the reference temperature and density gradients can be expressed via
the following thermodynamic relations
\begin{equation}
 \dfrac{\mathrm{d}\ln \tbar}{\mathrm{d}r} = \epsilon_S 
\dfrac{\mathrm{d}\sbar}{\mathrm{d}r}-Di\,\abar \gbar,
\label{eq:temp}
\end{equation}
and
\begin{equation}
 \dfrac{\mathrm{d}\ln \rbar}{\mathrm{d}r} = -Co\,\epsilon_S \abar \tbar 
\dfrac{\mathrm{d}\sbar}{\mathrm{d} r} - \dfrac{Di}{\Gamma_o} \dfrac{\abar 
\gbar}{\tilde{\Gamma}},
\label{eq:rho}
\end{equation}
where $\abar$ denotes the dimensionless expansion 
coefficient, while $\tilde{\Gamma}$ is the Gr\"uneisen parameter normalised by 
its value at $r_o$. The equations 
(\ref{eq:temp}-\ref{eq:rho}) involve four dimensionless parameters

\begin{equation}
 Di = \dfrac{\alpha_o g_o d}{c_p},\  Co=\alpha_o T_o, \ 
\Gamma_o, \  \epsilon_s = 
\dfrac{d}{c_p}\left|\dfrac{\mathrm{d}s}{\mathrm{d} r}\right|_{r_o}\,.
\end{equation}
According to the \textit{ab initio} calculations by \cite{French12}, 
the heat capacity $c_p$ exhibits little variation in most of Jupiter's interior 
and is hence assumed to be constant in the above equations.
$Di$ denotes the dissipation number, 
which characterises the ratio between the fluid layer thickness and the 
temperature scale, $Di=d/d_T$ with $d_T = c_p/\alpha_o g_o$. In the so-called 
\emph{thin-layer limit} of $d\ll d_T$, $Di$ vanishes and yields
the Boussinesq approximation of the Navier-Stokes equations
\cite[e.g.][]{Verhoeven15}. $Co$ is the compressibility number that is equal to 
unity when the fluid is an ideal gas, and is $\mathcal{O}(10^{-2})$
in liquid iron cores of terrestrial planets \citep[see][]{Anufriev05}. In the 
above 
equations, $\Gamma_o$ corresponds to 
the Gr\"uneisen parameter at the outer boundary, while 
$\epsilon_s$ characterises the departure of the 
background state from the adiabat. It has to satisfy $\epsilon_s \ll 1$ to 
ensure the consistency of the anelastic approximation \citep{Glatz1}.
In standard anelastic models such as the ones employed in the benchmarks 
by \cite{Jones11}, 
the background state is assumed to be a perfectly
adiabatic ideal gas (i.e. $\epsilon_s=0$, $Co=1$). 
The background state is in this case entirely specified by two parameters only: 
$Di$ and $\Gamma_o$, $Di$ being directly related to the number of 
density scale heights of the reference state \citep[see][their 
Eq.~2.9]{Jones09}, and $\Gamma_o$ is the inverse of the polytropic index.

At this stage, given that $\abar$ and $\tilde{\Gamma}$ directly depend on 
$\rbar$ and $\tbar$, the equations (\ref{eq:temp}-\ref{eq:rho}) coupled with 
the additional Poisson equation for gravity form a nonlinear problem that would 
necessitate an iterative solver \citep[for an example, see e.g.][]{Brun11}.
For the sake of simplicity and to ensure the future reproducibility of our 
results, we adopt here a grosser approach which consists of 
approximating $\abar$, $\gbar$ and $\tilde{\Gamma}$ by analytical functions 
which fit the interior model of \cite{French12}. The \ref{sec:ref_coeffs} 
enlists the numerical values of the approximated 
profiles of $\gbar$, $\abar$ and $\tilde{\Gamma}$.
A comparable approach was followed by \cite{Jones14} to define the reference 
state of his Jupiter dynamo models.

\begin{figure*}
 \centering
 \includegraphics[width=0.99\textwidth]{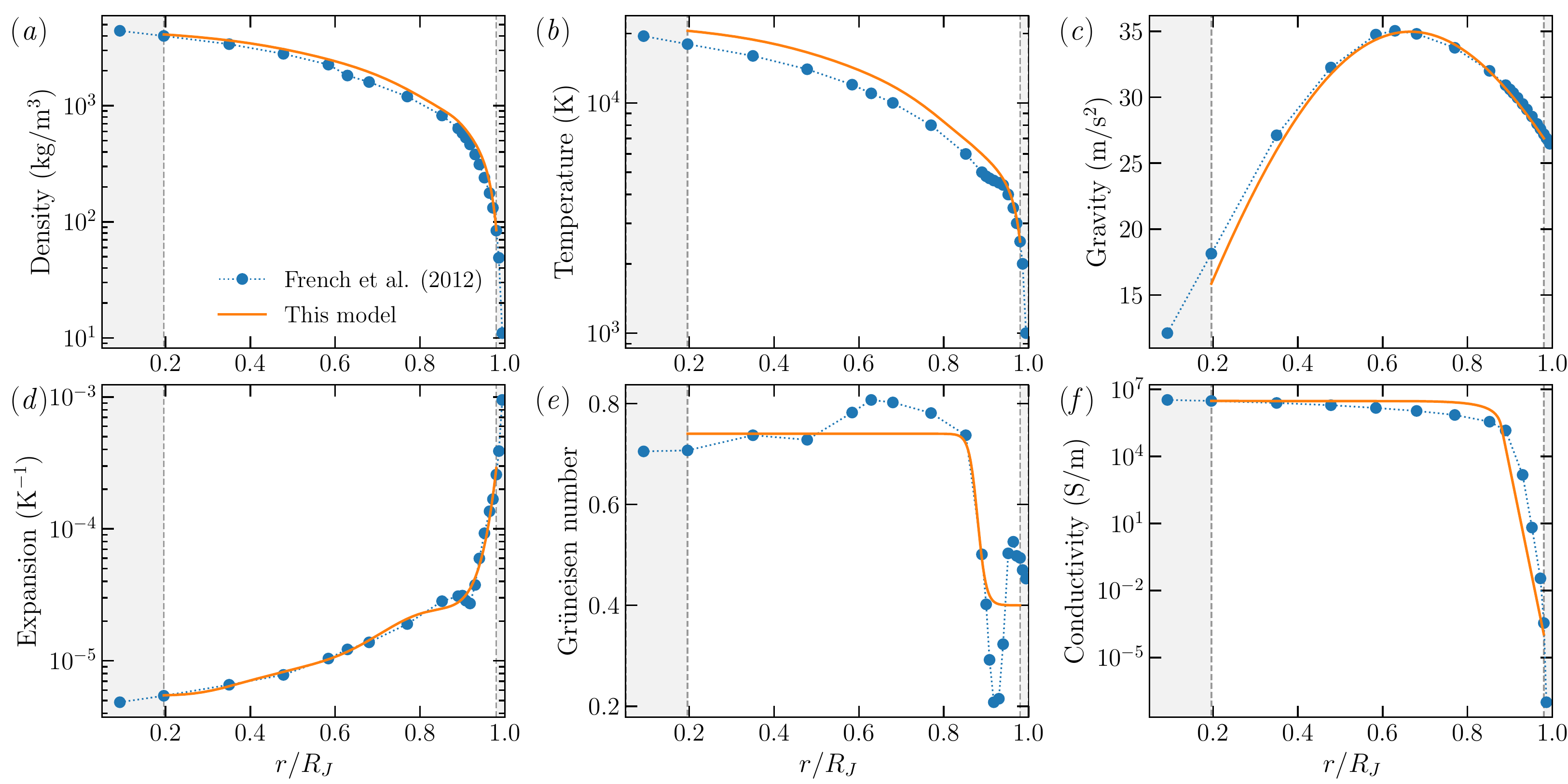}
 \caption{Comparison of the reference state considered in this study (solid 
lines in all panels) using $Di =28.417$, $Co=0.73$ and $\Gamma_o=0.4$ with the 
\textit{ab initio} models from \cite{French12} (dashed lines in all panels). 
(\textit{a}) Background density profile as a function of the normalised radius  
$r/R_J$. (\textit{b}) Background temperature profile as a function of the 
normalised radius. (\textit{c}) Gravity profile as a function of the normalised 
radius. (\textit{d}) Thermal expansion coefficient as a function of the 
normalised radius. (\textit{e}) Gr\"uneisen number as a function of the 
normalised radius. (\textit{f}) Electrical conductivity as a function of the 
normalised radius. (\textit{c}). The reference model employed in the numerical 
simulations spans from $0.196~R_J$ to $0.98~R_J$. Those boundaries are 
highlighted by gray shaded areas on each panel. The conversion between 
dimensional and dimensionless units is done by simple multiplication by the 
reference values expressed in Tab.~\ref{tab:ref}.}
\label{fig:profs}
\end{figure*}

Figure~\ref{fig:profs} shows a comparison between the reference state 
considered in this study using $Di =28.417$, $Co=0.73$ and $\Gamma_o=0.4$ 
(solid lines) with the \textit{ab initio} 
models from \cite{French12} (dashed lines). Most of the density and temperature 
contrasts are accommodated in the external $10\%$ of Jupiter's interior. Global 
models of rotating convection in anelastic spherical shells indicate that 
a steeply-decreasing background density goes along with smaller convective 
flow lengthscales \citep[e.g.][their Fig.~5]{Gastine12}. Resolving the entire 
density contrast up to the $1$~bar level would yield a lengthscale 
range that would become numerically prohibitive. As shown in 
Fig.~\ref{fig:profs}, we hence restrain the numerical fluid domain to an 
interval that spans $0.196\,R_J$ to $0.98\,R_J$, with $r_i/r_o=0.2$. Except 
explicitly-stated otherwise, the conversion between dimensionless and 
dimensional units is done by simple multiplication with the reference values at 
$r_o=0.98\,R_J$ given in Tab.~\ref{tab:ref}.

Though not fully thermodynamically consistent, the approximated reference 
state hence provides background profiles in good agreement with the 
interior models while keeping the reference state definition tractable.

\begin{table*}
  \centering
  \caption{Estimates of the physical properties of Jupiter's interior 
at two different 
depths. The material properties come from the \textit{ab initio} calculations 
from \cite{French12}. The magnetic field amplitude at $0.98\,R_J$ comes from 
\cite{Connerney18}, while the velocity and magnetic field estimates at depth 
come from the anelastic scaling laws by \cite{Yadav13a} and \cite{Gastine14a}.}
 \begin{tabular}{lrrr}
 \toprule
 Quantity & Notation &  \multicolumn{2}{c}{Value} \\
  \midrule
 Radius & $R_J$ & \multicolumn{2}{c}{$6.989\times 10^7$~m}\\
  Lengthscale & $d=0.8\times0.98 \times R_J$ & \multicolumn{2}{c}{$5.479\times 
10^7$~m}\\
  Rotation rate & $\Omega$ & \multicolumn{2}{c}{$1.75\times 10^{-4}$~s$^{-1}$}
\\
\midrule
  &  & Value at $0.196~R_J$ & Value at $0.98~R_J$ \\
  \midrule
 Density & $\rho$ & $3990$~kg$/$m$^{3}$  & $84.8$~kg$/$m$^{3}$\\
  Temperature & $T$ & $18000$~K & $2500$~K\\
  Gravity & $g$ & $18.1$~m$/$s$^{2}$ &$27.2$~m$/$s$^{2}$ \\
  Heat capacity & $c_p$ & $1.36\times 10^4$~J$/$kg$/$K & 
$1.29\times 
10^4$~J$/$kg$/$K\\
  Thermal expansion & $\alpha$ & $5.46\times 10^{-6}$~K$^{-1}$ & 
$2.58\times 10^{-4}$~K$^{-1}$ \\
Viscosity & $\nu$ & $2.66\times 10^{-7}$~m$^2/$s & $3.92\times 
10^{-7}$~m$^2/$s \\
Thermal diffusivity & $\kappa$ & $2.70\times 10^{-5}$~m$^2/$s &
$1.32\times 10^{-6}$~m$^2/$s\\
Electrical conductivity & $\sigma$ & $3.05\times 10^6$~S$/$m & 
$3.5\times 10^{-4}$~S$/$m \\    
Magnetic diffusivity & $\lambda$ & $0.261$~m$^2/$s &
$2.3\times 10^{9}$~m$^2/$s\\
Convective velocity & $u_c$ &$\mathcal{O}(10^{-2}-10^{-1})$~m$/$s & 
$1$~m$/$s 
\\
Zonal velocity & $u_Z$ &$\mathcal{O}(10^{-2}-10^{-1})$~m$/$s & 
$10$~m$/$s \\
Magnetic field strength & $B$ &$\mathcal{O}(10^{-2})$~T & $10^{-3}$~T \\
  \bottomrule
 \end{tabular}
 \label{tab:ref}
\end{table*}

\subsection{Transport properties}

The \textit{ab initio} calculations by \cite{French12} suggest that the 
kinematic viscosity is almost homogeneous in Jupiter's interior with 
values around $\nu\simeq 3\times 10^{-7}$~m$^2$/s (see 
Tab.~\ref{tab:ref}). In the following, we 
therefore simply adopt a constant kinematic viscosity. The thermal 
diffusivity exhibits a more complex variation. It gradually decreases
outward up to $0.9\,R_J$, above which it increases due to 
additional ionic transport becoming relevant there.
The overall variations are, however, limited to a factor of roughly 
$30$.
Following our previous models \citep{Gastine14a}, we neglect those 
variations and assume a constant thermal diffusivity $\kappa$ for 
simplicity. 
The electrical conductivity exhibits much steeper variations. A very abrupt 
increase inwards of the conductivity in the molecular envelope 
transitions around $0.9~R_J$ to shallower variations in the metallic core.
This profile is approximated in the numerical models by the continuous 
functions introduced by \cite{Gomez10}

\begin{equation}
 \tilde{\lambda}=\dfrac{1}{\tilde{\sigma}},\quad \tilde{\sigma} =\left\lbrace
 \begin{aligned}
  1+\left(\tilde{\sigma}_m-1\right)\left(\dfrac{r-r_i}{\hmag}\right)^{\xi_m}, 
\quad
r\leq r_m, \\
\tilde{\sigma}_m\exp\left(\xi_m\dfrac{r-r_m}{\hmag}\dfrac{\tilde{\sigma}_m-1}{
\tilde{\sigma}_m} \right), \quad r \geq r_m, \\
 \end{aligned}
 \right.
 \label{eq:cond}
\end{equation}
where $r_m$ is the radius that separates the two functions,
$\tilde{\sigma}_m$ denotes the dimensionless conductivity at $r_m$, $\xi_m$ the 
rate of the exponential decay and $\hmag=r_m-r_i$ is the thickness of the 
metallic region. Given the abrupt decay of electrical conductivity 
in the outer layer, we choose the value at the inner boundary $r_i$ for 
defining the reference magnetic diffusivity, in contrast with the other 
internal properties. 
Figure~\ref{fig:profs}\textit{f} shows a comparison between the electrical 
conductivity profile from \cite{French12} and Eq.~(\ref{eq:cond})
with the parameters $r_m=0.9\,r_o$, $\tilde{\sigma}_m=0.07$ and $\xi_m=11$ 
adopted in this study. The main difference between the two profiles arises in 
the metallic interior where we assume a constant electrical 
conductivity, while the \textit{ab initio} calculations suggest a 
linear increase with depth. While $Rm$ is limited to a few thousands in global 
models, it is expected to reach $\mathcal{O}(10^5-10^6)$ in Jupiter's interior 
\citep[e.g.][]{Yadav13a}.
We hence anticipate that the linear 
decrease of conductivity would have a much stronger dynamical impact at the 
moderate values of $Rm$ accessible to numerical dynamos than in 
Jupiter. Assuming a constant electrical conductivity in the lower layer at 
least guarantees that $Rm$ stays at a high level in this region.
To ensure that no spurious currents develop when the 
conductivity becomes too low at the external boundary, we assume that the 
electrical currents actually vanish when $\tilde{\sigma} < 10^{-5}$, i.e. when 
$r\geq 0.94\,r_o$ \citep[see][]{Elstner90,Dietrich18}.

\subsection{MHD equations}

Now that the spherically-symmetric and static background state and material 
properties have been specified, we consider the set of equations that govern the 
time evolution of 
the velocity $\vec{u}$, the magnetic field $\vec{B}$ and the entropy 
fluctuation $s'$. The equations are non-dimensionalised using 
the viscous diffusion time $d^2/\nu$ as the reference time scale, $\nu/d$ as 
the velocity unit and $\sqrt{\Omega \mu_0 \lambda_i \rho_o}$ as the reference 
scale for the magnetic field. The entropy fluctuations $s'$ are 
non-dimensionalised using the same unit as for $\sbar$, i.e. $d | \mathrm{d} 
s/\mathrm{d} r |_{r_o}$. This yields the following set of non-dimensional 
equations

\begin{equation}
 \vec{\nabla} \cdot (\rbar \vec{u}) = 0, \quad  \vec{\nabla} \cdot \vec{B} = 0,
\label{eq:soleno}
 \end{equation}
\begin{equation}
 \dfrac{D \vec{u}}{D
t}+\dfrac{2}{E}\vec{e_z}\times\vec{u} = -\vec{\nabla} 
\left(\dfrac{p'}{\rbar}\right)+\dfrac{1}{E 
Pm\,\rbar}\vec{j}\times\vec{B}-\dfrac{ 
Ra } { Pr } \abar\tbar \vec{g} s'+ \dfrac{1}{\rbar}\vec{\nabla}\cdot\tens{S},
\label{eq:NS}
\end{equation}
\begin{equation}
 \dfrac{\partial \vec{B}}{\partial t} = \vec{\nabla}\times 
\left(\vec{u}\times\vec{B}-\dfrac{\tilde{\lambda}}{Pm}\vec{\nabla}\times 
\vec{B} \right),
\label{eq:ind}
\end{equation}
and
\begin{equation}
 \rbar\tbar\left(\dfrac{D s'}{D 
t}+u_r\dfrac{\mathrm{d}\sbar}{\mathrm{d}r}\right) = 
\dfrac{1}{Pr}\vec{\nabla}\cdot\left(\rbar\tbar\vec{\nabla} s'\right)+\dfrac{Pr 
Di}{Ra}\left(\mathcal{Q}_\nu+\mathcal{Q}_\lambda\right),
\label{eq:heat}
\end{equation}
where $D/Dt=\partial/\partial t+\vec{u}\cdot\vec{\nabla}$ corresponds to the 
substantial time derivative, $p'$ is the pressure fluctuation, 
$\vec{j}=\vec{\nabla}\times\vec{B}$ is the current and $\tens{S}$ is the 
traceless rate-of-strain tensor expressed 
by
\begin{equation}
 \tens{S}_{ij} = 2\rbar\left(\tens{e}_{ij} - \dfrac{1}{3}\dfrac{\partial 
u_i}{\partial 
x_i}\right),\quad \tens{e}_{ij} = \dfrac{1}{2}\left(\dfrac{\partial 
u_i}{\partial x_j}+ 
\dfrac{\partial u_j}{\partial x_i}\right)\,.
\end{equation}
In Eq.~(\ref{eq:heat}), $\mathcal{Q}_\nu$ and $\mathcal{Q}_\lambda$ correspond 
to the viscous and Ohmic heating terms defined by

\begin{equation}
 \mathcal{Q}_\nu = 
2\rbar\left[\sum_{i,j}\tens{e}_{ij}\tens{e}_{ji}-\dfrac{1}{3}\left(\vec{\nabla}
\cdot\vec { u } \right)^2\right],\quad
 \mathcal{Q}_\lambda = \dfrac{\tilde{\lambda}}{E\,Pm^2}\vec{j}^2\,.
\end{equation}
Since global models cannot handle the small diffusivities of astrophysical 
bodies, we adopt here entropy diffusion as a primitive sub grid-scale model of 
thermal conduction \citep[see][]{Jones11}. This is a common approach in 
anelastic convective models \citep[see][]{Lantz99} which becomes more 
questionable when modelling the transition to stably-stratified layers. 
Comparison of numerical models with temperature and entropy diffusion by 
\cite{Lecoanet14} 
however yield quantitatively similar results. We hence adopt entropy diffusion 
throughout the entire fluid domain.

The set of equations (\ref{eq:soleno}-\ref{eq:heat}) is controlled by four 
dimensionless numbers, namely the Rayleigh number $Ra$, the Ekman number $E$, 
the Prandtl number $Pr$ and the magnetic Prandtl number $Pm$ 

\begin{equation}
 Ra=\dfrac{\alpha_o T_o g_o d^4 }{c_p \nu 
\kappa}\left|\dfrac{\mathrm{d} s}{\mathrm{d} r}\right|_o,\ E=\dfrac{\nu}{\Omega 
d^2},\ Pr=\dfrac{\nu}{\kappa},\ Pm=\dfrac{\nu}{\lambda_i}\,.
\end{equation}

For rapidly-rotating fluids, a relevant measure of the degree of stratification 
is the ratio of the Brunt-V\"ais\"al\"a frequency to the rotation rate 
\citep{Takehiro01}. This is related to the control parameter 
$\Gamma_s$ via

\begin{equation}
 \dfrac{N_m}{\Omega} = \max_r
\sqrt{\abar(r)\tbar(r)\gbar(r)\dfrac{Ra\,E^2}{Pr}\Gamma_s}\,.
\label{eq:N_m}
\end{equation}

\subsection{Boundary conditions}

We assume stress-free and impenetrable boundary conditions at both boundaries:
\begin{equation}
 u_r = \dfrac{\partial}{\partial r}\left(\dfrac{u_\theta}{r}\right)=
 \dfrac{\partial}{\partial r}\left(\dfrac{u_\phi}{r}\right)=0,\quad r=\lbrace 
r_i,r_o\rbrace\,.
\label{eq:bc_flow}
\end{equation}
Entropy is assumed to be fixed at the outer boundary, while the entropy 
gradient is imposed at the inner boundary:
\begin{equation}
 \left.\dfrac{\partial s'}{\partial r}\right|_{r=r_i}=0,\quad 
s'(r=r_o)=0\,.
 \label{eq:bc_ent}
\end{equation}
Fixing $s'$ at the outer boundary grossly reflects the entropy mixing in 
the neglected outer $2\%$ of Jupiter.
The material outside the simulated spherical shell is assumed to be 
electrically insulating. Hence, the magnetic field matches a potential 
field at both boundaries.

\subsection{Numerical methods}

The dynamo model presented in this study has been computed using the 
open-source MHD code \texttt{MagIC} \citep[freely available at 
\url{https://github.com/magic-sph/magic}, see][]{Wicht02}. \texttt{MagIC} has 
been tested and validated against several anelastic benchmarks \citep{Jones11}. 
The set of equations (\ref{eq:soleno}-\ref{eq:heat}) complemented by the 
boundary conditions (\ref{eq:bc_flow}-\ref{eq:bc_ent}) is solved in spherical 
coordinates by expanding the velocity and the magnetic fields into poloidal and 
toroidal potentials:

\begin{equation}
\begin{aligned}
 \rbar \vec{u} & =\vec{\nabla}\times(\vec{\nabla}\times 
W\,\vec{e_r})+\vec{\nabla} \times Z\,\vec{e_r}, \\
\vec{B} & =\vec{\nabla}\times(\vec{\nabla}\times G\,\vec{e_r})+\vec{\nabla} 
\times H\,\vec{e_r}\,.
\end{aligned}
\end{equation}
The quantities $W$, $Z$, $G$, $H$, $s'$ and $p'$ are expanded in spherical 
harmonics up to a degree $\ell_\text{max}$ in the angular directions and in 
Chebyshev polynomials up to the degree $N_c$ in the radial direction. For the 
latter, a Chebyshev collocation method is employed using the Gauss-Lobatto 
interval with $N_r$ grid points defined by
\[
 x_k = \cos\left[\dfrac{(k-1)\pi}{N_r-1}\right],\quad k\in[1,N_r]\,.
\]
This interval that ranges between $-1$ and $1$ is usually directly remapped 
onto $[r_i,r_o]$ by using a simple affine mapping 
\citep[e.g.][p.~468]{Glatz84}. However, because of the clustering of grid points 
in the vicinity of the boundaries, the Gauss-Lobatto grid features a minimum 
grid spacing that decays with $N_r^{-2}$. 
The propagation of Alfv\'en waves close to the boundaries then imposes severe 
restrictions on the time step size \citep{Christensen99}. To alleviate this 
limitation, we rather employ the mapping by \cite{Kosloff93} defined by
\[
 r_k = \dfrac{r_o-r_i}{2}\dfrac{\arcsin(\alpha_{\text{map}} x_k)}{\arcsin 
\alpha_\text{map}}+\dfrac{r_o+r_i}{2}, \quad k\in[1,N_r]\,.
\]
To ensure the spectral convergence of the collocation method, the mapping 
coefficient $\alpha_\text{map}$ has to be kept under a maximum value that 
depends on $N_r$
\[
 \alpha_\text{map} \leq \left[\cosh\left(\dfrac{|\ln 
\epsilon_m|}{N_r-1}\right)\right]^{-1},
\]
where $\epsilon_m$ is the machine precision \citep{Kosloff93}.

The equations are advanced in time using an implicit-explicit 
Crank-Nicolson Adams-Bashforth second order scheme, which handles the nonlinear 
terms and the Coriolis force explicitly and the remaining terms implicitly 
\citep{Glatz84}. Because of the stable stratification, the advection of the 
background entropy gradient, $u_r \mathrm{d}\sbar /\mathrm{d} r$, that enters 
Eq.~(\ref{eq:heat}) is also handled implicitly to avoid severe time step 
restrictions when the Brunt-V\"ais\"al\"a frequency exceeds the rotation rate 
\citep[see][]{Brown12}. \texttt{MagIC} uses the open-source library 
\texttt{SHTns} \citep[freely available at 
\url{https://bitbucket.org/nschaeff/shtns}, see][]{Schaeffer13} for the 
spherical harmonic transforms. A more comprehensive description of the 
numerical method can be found in \cite{Glatz84}, \cite{Tilgner99} or 
\cite{Christensen15}.

\begin{table*}
 \centering
 \caption{Definitions and estimates of dimensionless parameters in Jupiter's 
interior along with values adopted in the numerical model. Estimates 
for Jupiter have been obtained using the dimensional values from 
Tab.~\ref{tab:ref}. The deviation from 
the adiabat $\epsilon_s$ has been obtained by using a simple thermal wind 
balance $\epsilon_s \sim \Omega\, u / \alpha_o g_o T_o$ \citep[see][]{Jones15}.
The estimates of the degree of stratification and the location of a possible 
SSL in Jupiter come from \cite{Militzer16} and \cite{Debras19}. The 
mean density $\rho_m=1300$~kg$/$m$^3$ and the mean magnetic
diffusivity $\lambda_m = 1.15$~m$^2/$s come from \cite{French12}.}
 \begin{tabular}{lllrr}
 \toprule
 Symbol & Name & Definition & Jupiter & This model \\
  \midrule
  $Di$ & Dissipation & $\alpha_o T_o g_o/c_p$ & $29.8$ & $28.42$  \\
  $Co$ & Compressibility &$\alpha_o T_o$ &$0.645$ & $0.73$ \\
  $\Gamma_o$& Gr\"uneisen && $0.470$ & $0.4$   \\
  $\epsilon_s$ & Adiabaticity & $d\,|\mathrm{d}s/\mathrm{d}r|_{r_o}/c_p$ & 
$\mathcal{O}(10^{-6})$ & $10^{-4}$  \\
$\rins$ & SSL inner radius & &$0.8-0.9\,R_J$ &  $0.82\,R_J$  \\ 
$\routs$ & SSL outer radius & &$0.88-0.93\,R_J$ & $0.86\,R_J$  \\
  $N_m/\Omega$ & Degree of stratification & & $1-3$ & $10.4$  \\
  \midrule
  $Ra$  & Rayleigh & $\alpha_o T_o g_o d^4 | \mathrm{d} s 
/\mathrm{d} r|_{r_o}/\nu\kappa c_p$ & $10^{31}$ & $3.7\times 10^{10}$ \\
  $E$ & Ekman & $\nu /\Omega\,d^2$ & $10^{-18}$ & $10^{-6}$  \\
  $Pr$& Prandtl & $\nu / \kappa$ & $10^{-2}-1$ & $0.2$ \\
  $Pm$ & Magnetic Prandtl &$\nu/\lambda_i$ & $10^{-6}$ & $0.4$  \\
  \midrule
  $Rm$ &  Magnetic Reynolds &$ u\,d /\lambda_i$ & $\mathcal{O}(10^6)$ & 
$4.11\times 10^2$ \\
  $Re$ & Reynolds &$ u\,d /\nu$ & $\mathcal{O}(10^{12})$ & $6.23\times 10^3$  
\\
  $Ro$ & Rossby & $ u / \Omega\,d$ & $\mathcal{O}(10^{-6})$ & $6.23\times 
10^{-3}$ 
\\
  $Re_Z$ & Zonal Reynolds &$ u_z\,d/\nu$ & $\mathcal{O}(10^{12}-10^{15})$ & 
$5.79\times 10^3$ \\
  $Re_c$ & Convective Reynolds &$ u_c\,d/\nu$ & $\mathcal{O}(10^{12})$ & $2.33 
\times 10^3$ \\
  $\Lambda$ & Elsasser & $ B^2 / \rho_m \lambda_m \mu_0 \Omega$ & 
$\mathcal{O}(10^{1}-10^{2})$ & $8.52$ 
\\
  $\overline{E_M}/\overline{E_K}$ & Energy ratio & $B^2 / \mu_0 \rho_m u^2 $ 
& $\mathcal{O}(10^2-10^3)$ & $3.50$ \\
  $f_{ohm}$ & Ohmic fraction 
&$\overline{\mathcal{D}_\lambda}/(\overline{\mathcal{D}_\lambda}
+\overline{\mathcal{D}_\nu})$ &  $1$ & $0.78$\\
  $f_{dip}$ & Axial-dipole fraction & $B^2_{\ell=1,m=0} (R_J)/ 
B^2_{\ell,m\leq 
12}(R_J)$  & $0.75$ & $0.95$  \\
  \bottomrule
 \end{tabular}
 \label{tab:params}
\end{table*}

\subsection{Control parameters}

The formation of zonal flows in global spherical models requires 
a combination of strong turbulent convective motions (i.e. large Reynolds 
numbers) and rapid rotation (i.e. low Rossby numbers). This 
regime, frequently referred to as the \emph{quasi-geostrophic turbulent regime} 
of 
convection \citep[e.g.][]{Julien12a}, can only be reached a low enough Ekman 
numbers, where global numerical simulations become 
extremely demanding. We therefore focus here on one single global 
dynamo model with $E=10^{-6}$, $Ra=3.7\times 10^{10}$, 
$Pm=0.4$, $Pr=0.2$. We adopted a spatial resolution 
of $N_r=361$ (with $\alpha_\text{map}=0.994$)  and  $\ell_\text{max}=597$ for 
most of the run. For the alias-free mapping used in the horizontal directions, 
this corresponds to $N_\theta=896$ latitudinal grid points and
$N_\phi=1792$ longitudinal grid points.
Spatial convergence of the solution has been tested by 
increasing the angular resolution to $\ell_\text{max}=1024$ ($N_\phi=3072$) 
towards the end of the run, without any noticeable change in the average 
properties. 
To ease the transients, the numerical model was initiated from 
another dynamo simulation computed at a larger Ekman number, and mild 
hyper-diffusion of the velocity and entropy fields were used over the first half 
of the computation time before their gradual removal \citep[e.g.][]{Kuang99}.

In rapidly-rotating convection \citep[e.g.][]{Takehiro01,Dietrich18a,Gastine20}, 
the distance of penetration $\delta$ of a convective eddy of size $d_c$ is 
directly related to the ratio of the Brunt-V\"ais\"al\"a frequency to the 
rotation rate via
\begin{equation}
 \delta = \left(\dfrac{N_m}{\Omega}\right)^{-1} d_c\,.
 \label{eq:penet}
\end{equation}
Ensuring that $\delta$ remains smaller than the thickness of the SSL 
$\hstrat$ requires $N_m/\Omega > 
d_c/\hstrat$. A thinner layer would thus require a stronger stratification or 
a slower rotation to remain effective. Here we adopt $\hstrat=0.05$ and 
$N_m/\Omega\simeq 10.4$ ($\Gamma_s=2000$).
The strong degree of stratification should suffice to stop even very large 
eddies of half the system size, $d_c\simeq 0.5\,d$. 
While thinner and shallower stable layers may be compatible with gravity 
observations, they would also considerably increase the numerical costs.  
Increasingly fine spatial grids are required to resolve the dynamics of 
thinner 
layers. Moreover, the tendency to form multiple jets increases with decreasing 
Ekman number. Relevant here is the effective Ekman number of the 
outer layer $E_o=E (d/d_o)^2$ with
$d_o=r_o-\routs$. Multiple jets may start to form below $E_o\approx 10^{-4}$ 
\citep[e.g.][]{Jones09,Gastine14}, a value barely reached for $d_o=0.12$ and 
$E=10^{-6}$.

The upper parts of Tab.~\ref{tab:params} summarises our control parameters as 
well as the corresponding values for Jupiter. 

Because of its significant numerical cost, the dynamo model has been integrated 
for a bit more than $0.13$ magnetic diffusion time (or $8400$ rotation periods), 
which required roughly $3$ million core hours on Intel Haswell CPUs.

\subsection{Diagnostics}

We analyse the numerical solution by defining several diagnostic properties. 
In the following, triangular brackets denote volume 
averaging, square brackets azimuthal averaging and overlines 
time averaging

\[
 \langle f\rangle = \dfrac{1}{V}\int_V f\,\mathrm{d}V,\ [ 
f ] = \dfrac{1}{2\pi}\int_0^{2\pi} f\,\mathrm{d}\phi,\ 
 \bar{f} = \dfrac{1}{\tau}\int_{t_o}^{t_o+\tau} f\,\mathrm{d} 
t\,,
\]
where $\tau$ is the averaging interval and $V$ is the spherical shell volume. 
Since the background state strongly varies with radius, it is also convenient 
to explore averages over a spherical surface 
\[
 \| f \|(r,t)=\int_{0}^{2\pi}\int_{0}^{\pi} |f| 
\sin\theta\,\mathrm{d}\theta\,\mathrm{d}\phi\,.
\]
%
%

The typical convective flow amplitude is measured by the Reynolds number $Re$, 
the Rossby number $Ro$ or the magnetic Reynolds number $Rm$ defined by

\begin{equation}
 Re = \sqrt{\overline{\langle \vec{u}^2 \rangle}},\quad Ro = Re\,E,\quad Rm 
=\overline{\dfrac{1}{V}\int_{r_i}^{r_o} \dfrac{\sqrt{\| \vec{u}^2 
\|}}{\tilde{\lambda}} r^2 \mathrm{d} r}\,.
\label{eq:vel_measure}
\end{equation}
To better separate the different flow components, we define two 
additional measures based on the zonal flow velocity, $Re_z$, and on the 
convective flow velocity, $Re_c$:

\begin{equation}
 Re_z = \sqrt{\overline{ \langle [u_\phi]^2 \rangle}},\quad Re_c = 
\sqrt{Re^2-Re_z^2}\,.
\end{equation}
The magnetic field amplitude is characterised by the Elsasser number 
\begin{equation}
 \Lambda = \left\langle \dfrac{B^2}{\tilde{\rho}\tilde{\lambda}}\right\rangle\,.
\end{equation}
The geometry of the surface magnetic field is expressed by its 
axial dipolar 
fraction $f_{dip}$, which is defined as the ratio of the energy of the 
axisymmetric dipole component to the magnetic energy in the spherical harmonic 
degrees $\ell \leq 12$ at $r_o$ \citep{Christensen06}.

The numerical solution is also examined in terms of its power budget.
Taking the inner product of the Navier-Stokes equation (\ref{eq:NS}) 
by $\vec{u}$ and the 
induction equation (\ref{eq:ind}) by $\vec{B}$ yields
\begin{equation}
 \dfrac{d}{dt}\left(E_K +E_M\right) =
\mathcal{P}-\mathcal{D}_\nu-\mathcal{D}_\lambda\,.
\label{eq:power_bal}
\end{equation}
In the above equation, $E_K$ and $E_M$ denote the mean kinetic and magnetic 
energy densities

\[
 E_K(t) = \dfrac{1}{2}\left \langle \tilde{\rho} \vec{u}^2 \right \rangle, 
\quad
 E_M(t) = \dfrac{1}{2}\dfrac{1}{E\,Pm}\left \langle \vec{B}^2 \right \rangle,
\]
$\mathcal{P}$ is the buoyancy power density
\[
 \mathcal{P}(t)= \dfrac{Ra E}{Pr} \left \langle 
\tilde{\alpha}\tilde{T} \tilde{g} s' u_r \right \rangle\,,
\]
and $\mathcal{D}_\nu$ and $\mathcal{D}_\lambda$ the power dissipated by viscous 
and Ohmic effects

\[
\mathcal{D}_\nu(t) = \langle \tens{S}^2 \rangle,\quad
 \mathcal{D}_\lambda(t) = \dfrac{1}{E Pm^2}\langle \tilde{\lambda} \vec{j}^2 
\rangle\,.
\]
Once a statistically-steady state has been reached, time averaging
Eq.~(\ref{eq:power_bal}) yields a balance between buoyancy input
power and heat losses by Ohmic and viscous dissipations
\begin{equation} 
\overline{\mathcal{P}}-\overline{\mathcal{D}_\nu}-\overline{\mathcal{D}_\lambda}
 = \overline{\mathcal{P}} 
-\dfrac{1}{f_{ohm}}\overline{\mathcal{D}_\lambda}= 0\,,
\end{equation}
where $f_{ohm}$ quantifies the fraction of heat dissipated ohmicly.
The residual in the above equation 
can serve as a good indicator of the time and spatial convergence of a
numerical solution \citep[e.g.][their Fig.~2]{King12}. 
Here, this identity is obtained to a high degree of fidelity with
$|\overline{\mathcal{P}}-\overline{\mathcal{D}_\nu}-\overline{\mathcal{D}
_\lambda}| / \overline{\mathcal{P}} < 0.3\%$.

Table~\ref{tab:params} summarises the control parameters and the main 
diagnostics of the dynamo model presented here along with the expected values 
for Jupiter. For comparison we note that the Jovian dynamo model by 
\cite{Gastine14a} was computed using $Pm=0.6$ and $E=10^{-5}$. It produced a 
relatively weak-field solution with $\overline{E_M} /\overline{E_K} \simeq 
0.1$ and $f_{ohm}\simeq 0.15$. In contrast, by employing much larger 
magnetic Prandtl number ($Pm \geq 3$), several dynamo simulations by 
\cite{Jones14} and \cite{Duarte18} yielded a stronger magnetic field with 
$\overline{E_M}/\overline{E_K} \simeq 3$. Adopting a significantly lower Ekman 
number enables us to reach a comparable energy fraction 
$\overline{E_M}/\overline{E_K} \simeq 3.5$ while using a magnetic 
Prandtl number almost one order of magnitude smaller. This yields an Ohmic 
fraction $f_{ohm} \simeq 0.8$, much closer to the value expected for 
Jupiter where Ohmic dissipation dominates by far because of the 
small magnetic Prandtl number.
Using $E=10^{-6}$ also ensures that $Re \gg 1 $ and yet $Ro 
\ll 1$, two prerequisites to develop turbulent quasi-geostophic convection 
conducive for sustaining strong zonal jets.

\section{Results}

\label{sec:results}

\subsection{Convective flow and magnetic field morphology}

\begin{figure*}
 \centering
 \includegraphics[width=.95\textwidth]{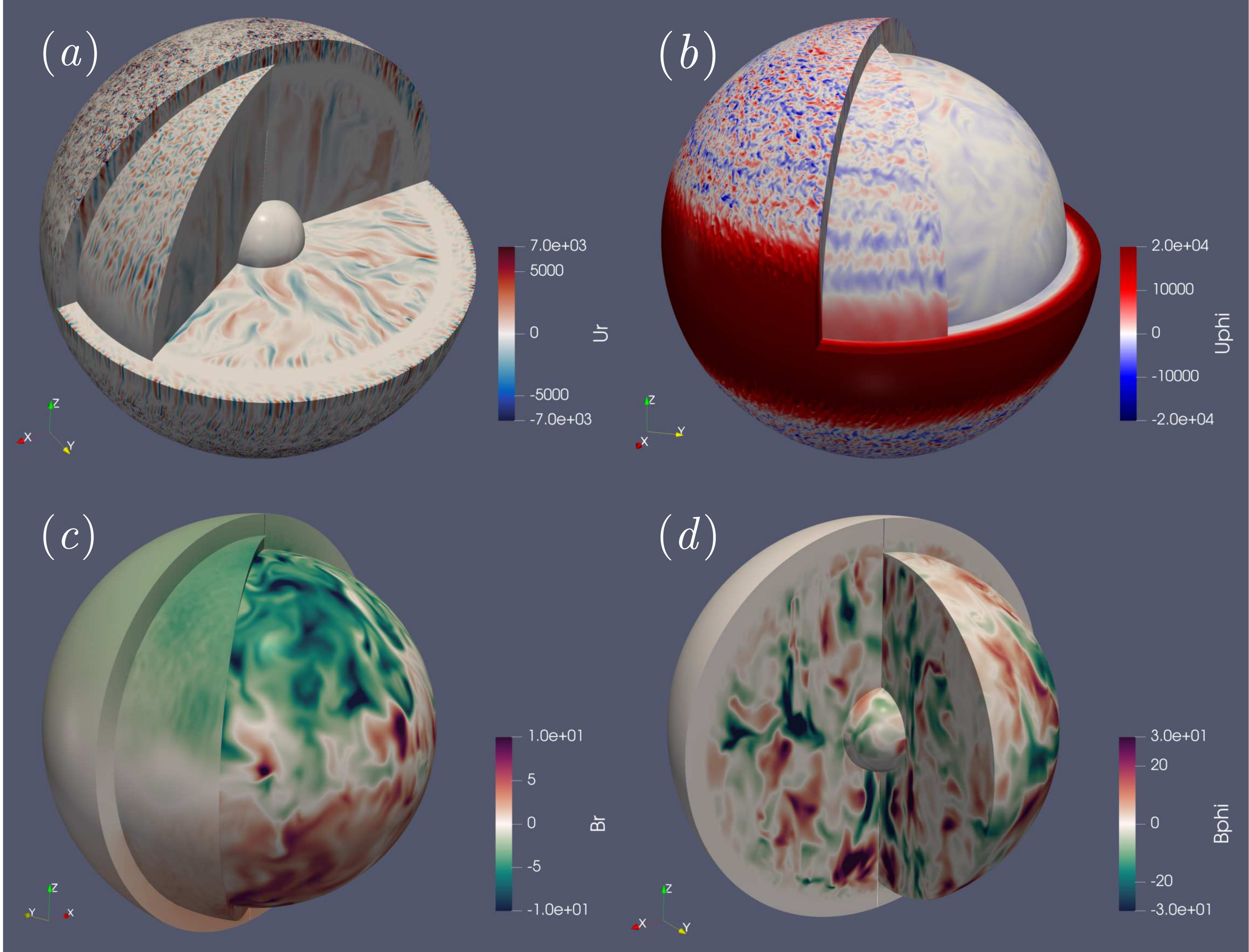}
 \caption{ 3-D renderings of the radial velocity $u_r$ (\textit{a}), of the 
azimuthal velocity $u_\phi$ (\textit{b}), of the 
radial component of the magnetic field $B_r$ (\textit{c}) and of the azimuhtal 
component of the magnetic field $B_\phi$ (\textit{d}). The inner sphere in 
panels (\textit{a}) and (\textit{d}) is located very close to the inner boundary 
at $r=r_i+0.01$, while in panel (\textit{b}) and (\textit{c}) it depicts the 
lower boundary of the SSL.
The intermediate radial cut that spans $30^\circ$ in longitude 
in panels (\textit{a})-(\textit{c}) and $90^\circ$ in panel (\textit{d}) is 
located close to the upper boundary of the stably-stratified layer at 
$r=0.904\,r_o$. The external radial cut corresponds to $r=0.992\,r_o$ in panel 
(\textit{a}) and to the surface $r_o$ in the other panels.}
 \label{fig:snap}
\end{figure*}


We start by examining the typical convective flow and magnetic field 
produced by the numerical dynamo model. Figure~\ref{fig:snap} shows a selected 
snapshot of the radial and azimuthal components of the velocity and magnetic 
fields.
An immediate effect of the strong stratification $N_m/\Omega \simeq 10$ is to 
significantly inhibit the convective motions between $\rins$ and $\routs$. 
The equatorial and meridional cuts of the radial velocity 
(Fig.~\ref{fig:snap}\textit{a}) clearly show that the SSL forms a 
strong dynamical 
barrier between two different convective regions. In the deep interior, the 
convective pattern takes the form of radially-elongated quasi-geostrophic 
sheets that 
span most of the metallic core. This is a typical flow pattern commonly 
observed in geodynamo models when the magnetic energy 
exceeds the kinetic energy \citep[e.g.][their Fig.~2]{Yadav16}.
In contrast, the outer layer is dominated by small-scale turbulent features. 
Because of the rapid decrease of density there, the convective flow become 
smaller-scale and more turbulent towards the surface. The azimuthal 
flows are dominated by a strong prograde equatorial jet which penetrates down 
to
$\routs$ (Fig.~\ref{fig:snap}\textit{b}) but are then effectively quenched in 
the stable layer. Flanking weaker jets of alternating direction appear up to 
about $\pm40^\circ$ in latitude. They become somewhat more pronounced with depth 
and show clearer at $r=0.9\,r_o$ ((radial cut in Fig.~\ref{fig:snap}\textit{b})

The magnetic field is predominantly produced in the metallic region below 
$\rins$ where both the 
conductivity and the convective flow amplitude are sufficient to sustain dynamo 
action (Fig.~\ref{fig:snap}\textit{c}-\textit{d}).
The magnetic Reynolds number (Eq.~\ref{eq:vel_measure}) reaches values 
of more than $600$ in this region.
The magnetic field at the top of the inner convective region features a 
dominant axisymmetric dipole accompanied by intense localised flux 
patches (inner radial cut in Fig.~\ref{fig:snap}\textit{c}).
Because of the strong inhibition of the flow motions between $\rins$ and 
$\routs$, there is little to no dynamo action happening in the SSL. Instead, 
the SSL filters out the faster varying field components via a magnetic 
skin effect as will be discussed in the next section
\citep[e.g.][]{Christensen06a,Gastine20}. Since smaller scale 
contributions vary  on shorter time scales, the remaining field at $\routs$ is 
of much larger scale than at $\rins$.

Because of the abrupt drop of electrical conductivity in the molecular 
envelope, the dynamo action in the outer layer is very inefficient since 
the magnetic Reynolds number $Rm$ is mostly smaller than one in the 
outer convective layer of our simulation.
Consequently, the locally-induced poloidal field remains practically 
negligible and the 
magnetic field decays like a potential field  with radius \citep{Wicht19}. 
The surface magnetic field is dominated by a strong axial dipole combined with 
large scale non-axisymmetric flux patches. 
In the fully-convective models by \cite{Gastine14a},  
the prograde equatorial jet shears the upper layers of the metallic 
region to produce strong azimuthal magnetic bands \citep{Wicht19}. Such 
structures are not observed here (Fig.~\ref{fig:snap}\textit{d}), likely because 
the zonal motions are hampered in the stable layer.

\subsection{Energetics}

\begin{figure*}
 \centering
  \includegraphics[width=16cm]{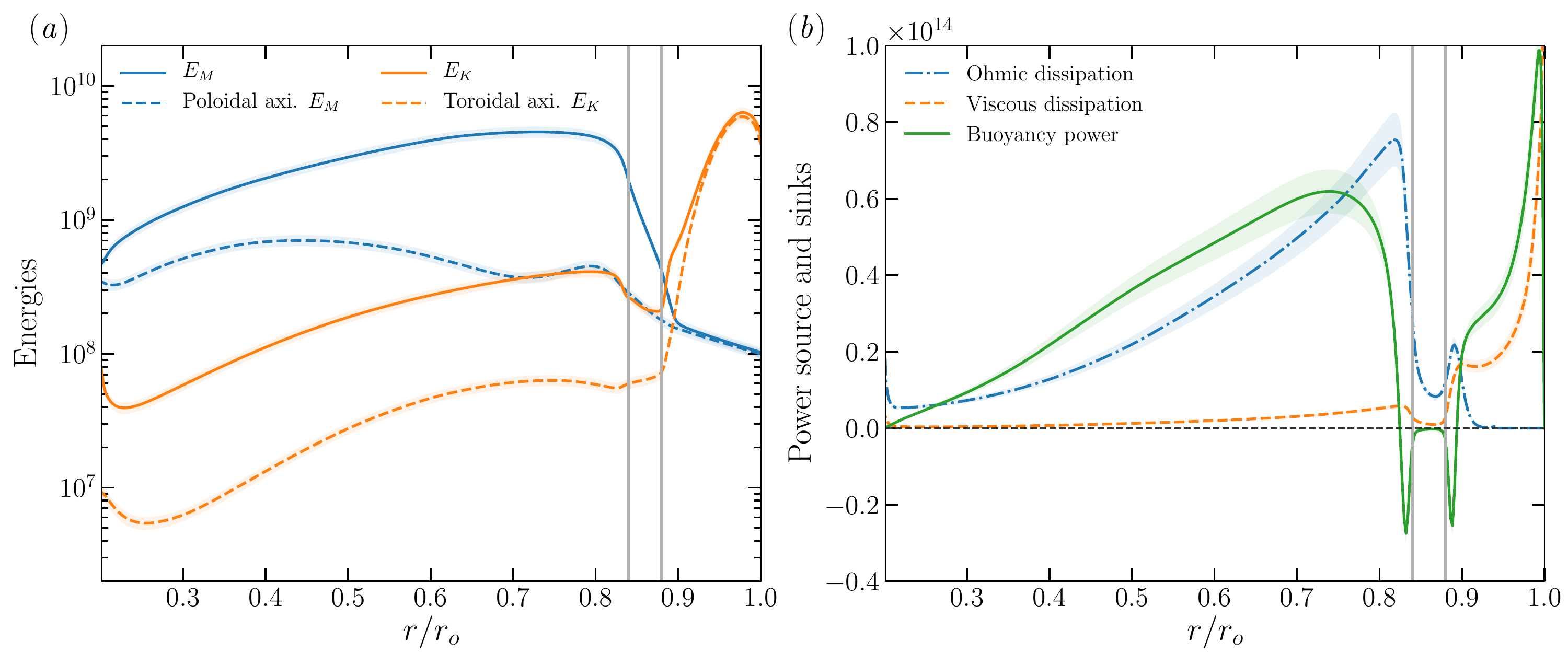}
  \caption{(\textit{a}) Time-averaged radial profiles of magnetic and kinetic 
energies. (\textit{b}) Time-averaged radial profiles of Ohmic and viscous 
dissipation and buoyancy power. The shaded area correspond 
to one standard-deviation accross the mean. The vertical lines mark the 
location of the stably-stratified layer between $\rins$ and $\routs$ 
(see Fig.~\ref{fig:dsdr}).}
  \label{fig:radprofs}
\end{figure*}

For a more quantitative assessment, we now examine the power balance.
Figure~\ref{fig:radprofs} shows the time-averaged radial profiles of magnetic 
and kinetic energy as well as the different source and sinks which enter the 
power balance (\ref{eq:power_bal}). In the metallic interior, the total 
magnetic energy exceeds the 
kinetic energy by one order of magnitude and zonal winds (toroidal axisymmetric) 
contribute only about $10$\% of the kinetic energy 
(Fig.~\ref{fig:radprofs}\textit{a}).
In the SSL, the total kinetic energy drops with radius by about a factor 
of two. 
Non-axisymmetric contributions drop more rapidly, but this is partly 
compensated by an increase in the zonal kinetic energy due to 
penetation from the upper convective region. In 
the external convective layer, fast zonal winds clearly dominate and the 
kinetic energy reaches its peak value at about $0.98\,r_o$.
Because of the skin-effect in the SSL and the decay 
of conductivity, the magnetic field becomes more axisymmetric and poloidal 
towards the surface. 
While kinetic and magnetic energy reach a comparable level at the top of 
the SSL, the former exceeds the latter by up to a factor $50$ in the outer 
convective layer. 

The sign changes of the buoyancy power (Fig.~\ref{fig:radprofs}\textit{b}) mark 
the actual separation between the convective and the stably-stratified layers. 
In the convective regions, the eddies which carry a positive entropy 
fluctuation compared to their surroundings ($s'>0$) rise outwards, while the 
ones with $s'<0$ sink inwards, yielding a positive correlation between $u_r$ and 
$s'$ and hence a positive buoyancy power. The opposite happens when a 
convective feature overshoots in an adjacent sub-adiabatic region. A rising 
parcel of fluid with $u_r>0$ now carries a perturbation $s'<0$ (and hence 
$\mathcal{P} < 0$) until it is homogenised with its surroundings by heat 
conduction ($\mathcal P \simeq 0$). Because of the finite 
stiffness of the background entropy gradient (Fig.~\ref{fig:dsdr}), 
the actual thickness of the region with $\mathcal{P}<0$ exceeds the interval 
$[\rins, \routs]$ delineated by vertical lines in Fig.~\ref{fig:radprofs}.
The measure of the vertical extent of the 
regions with $\mathcal{P}<0$ actually provide a good estimate of the distance of 
penetration of the convective eddies into a stably-stratified layer 
\citep[e.g.][]{Browning04,Takehiro18a,Gastine20}. 

In line with the partitioning between magnetic and kinetic energies, the heat 
losses are dominated by Ohmic heating in the metallic region, while viscous 
heating takes over when the electrical conductivity drops.
To sum up, Fig.~\ref{fig:radprofs} highlights the separation between two 
different dynamical regions: an internal metallic region, which harbours the 
production of a strong magnetic field, and an external envelope where most of 
the kinetic energy is pumped into zonal motions.

\begin{figure*}
 \centering
 \includegraphics[width=16cm]{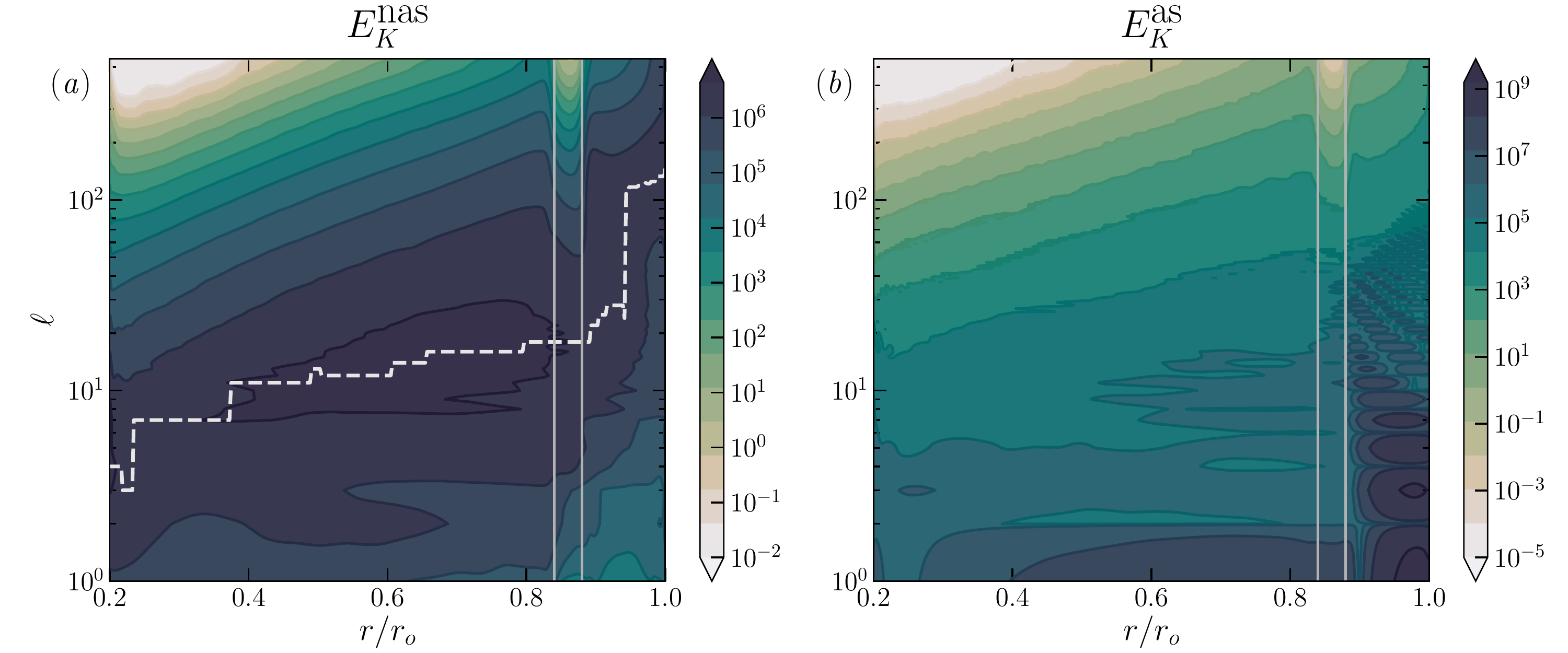}
  \includegraphics[width=16cm]{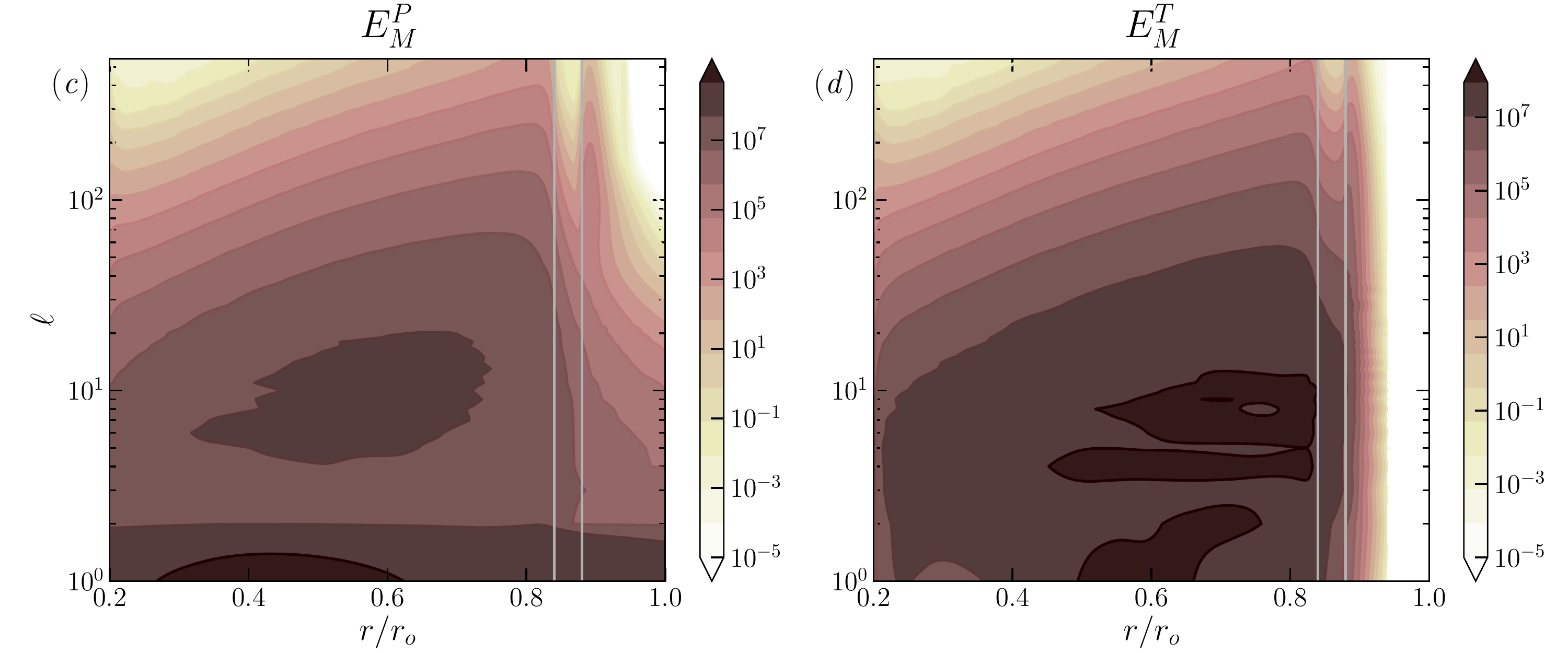}
 \caption{Time-averaged 2-D spectra in $(r/r_o,\ell)$ plane for several kinetic 
(upper panels) and magnetic (lower panels) contributions: (\textit{a}) 
non-axisymmetric kinetic energy, (\textit{b}) axisymmetric kinetic 
energy, (\textit{c}) poloidal magnetic energy, and (\textit{d}) toroidal 
magnetic energy. The thick dashed line in panel (\textit{a}) mark the location 
of the maxima of non-axisymmetric kinetic energy. The solid lines mark the 
location of the stably-stratified layer between $\rins$ and $\routs$ 
(see Fig.~\ref{fig:dsdr}). Because of the different dynamics, the colorbars are 
different for each panel.}
 \label{fig:spec_l_r}
\end{figure*}

To better characterise the dynamics in the different layers, we now examine the 
spectral energy distributions in the $(r/r_o,\ell)$ plane.
Figure~\ref{fig:spec_l_r} illustrates 2-D spectra of kinetic (upper panels) and 
magnetic (lower panels) energy contributions. For a 
more insightful analysis, the kinetic energy has been split into 
non-axisymmetric (Fig.~\ref{fig:spec_l_r}\textit{a}) and axisymmetric 
(Fig.~\ref{fig:spec_l_r}\textit{b}) motions, while the magnetic spectra 
have been separated into poloidal (Fig.~\ref{fig:spec_l_r}\textit{c}) and 
toroidal (Fig.~\ref{fig:spec_l_r}\textit{d}) contributions.

We introduce the local peak of the 
non-axisymmetric energy $\hat{\ell}$ and the corresponding convective flow
lengthscale $d_c$ defined by
\begin{equation}
 \hat{\ell}(r) = \argmax_\ell E_K^{\text{nas}}, \quad 
d_c(r)=\dfrac{\pi\,r}{\hat{\ell}},
\label{eq:hat}
\end{equation}
where $E_K^{\text{nas}}$ denotes the non-axisymmetric energy 
\citep[e.g.][]{Schwaiger19}.
The kinetic energy spectra clearly differ for the three regions. 
In the external layers ($r>\routs$), the convective 
lengthscale rapidly decreases outwards, reaching $\hat{\ell}\sim 100$, i.e. 
$d_c\approx 0.03\,r_o$. The scale of the zonal flows remains roughly an order 
of magnitude larger with $\ell\le20$. 
In the metallic core
($r<\rins$), $\hat{\ell}$ decreases only mildly from about $15$ at $\rins$ to 
about $4$ at $r_i$. 
In the physical space this corresponds to the large 
scale convective sheets visible in Fig.~\ref{fig:snap}\textit{a}. In between 
those two regions, 
the stably-stratified layer significantly reduces the amplitude of the 
convective motions. The inhibition of the convective flow depends on 
the size of the convective eddies: the smaller the lengthscale, the stronger 
the attenuation of the kinetic energy. This phenomenon can be understood when 
considering the distance of penetration $\delta$ of a turbulent feature of 
horizontal size $d_c$ into a stably stratified layer (Eq.~\ref{eq:penet}).
Approximating the horizontal scale $d_c$ by $\pi \rins/\ell$ then yields

\begin{equation}
 \delta_\ell \sim \dfrac{\pi}{\rins}\left(\dfrac{N_m 
\ell}{\Omega}\right)^{-1}\,.
\end{equation}
The penetration distance $\delta_\ell$ is hence inversely proportional to the 
degree $\ell$, explaining the stronger damping of small convective scales 
\citep{Dietrich18a}.
At the dominant lengthscale of convection $\hat{\ell}\simeq 20$ at the edges of 
the SSL, the above scaling yields $\delta_{\hat{\ell}} \simeq 0.015\,d$, in 
good agreement with the actual thickness of the overshoot regions characterised 
by $\mathcal{P}<0$  (Fig.~\ref{fig:profs}\textit{b}).

The poloidal magnetic energy is dominated by its dipolar component throughout 
the entire volume. In the metallic interior, it features a secondary peak 
around $\ell \simeq 10-20$ which roughly follows the variations of the peak of 
the non-axisymmetric kinetic energy $\hat{\ell}(r)$ \citep{Aubert17}. 
 The toroidal field reaches its maximum amplitude in the upper half of the 
metallic core ($0.6 \leq r \leq \rins$) and also peaks at comparable scales.
Beyond $\rins$, the magnetic energy decreases up to the surface $r_o$ and is 
significantly more 
attenuated at small scales. This phenomenon arises because of two distinct 
scale-dependent physical processes:

\begin{enumerate}
 \item Within the 
SSL, the electrical conductivity is almost as large as in the metallic core but 
the convective motions are significantly hampered. A first order approximation 
assumes that the SSL behaves as a 
stagnant layer of size $\hstrat$ with a constant electrical conductivity. Such 
a layer will attenuate the poloidal magnetic 
energy by \emph{skin effect} \citep[e.g.][]{Christensen06a} by a factor

\begin{equation}
 \ln\dfrac{E_{M}^{P,\ell} (\routs)}{E_{M}^{P,\ell}(\rins)} \sim 
-\dfrac{\hstrat}{\delta_\ell^{\text{SK}}},
\label{eq:skin}
\end{equation}
where $E_{M}^{P,\ell}$ is the poloidal magnetic energy at the 
harmonic degree $\ell$ and 
$\delta_\ell^{\text{SK}}$ is the skin depth associated with a 
feature of scale $d_c$ expressed by \citep[see][]{Gastine20}
\begin{equation}
 \delta_\ell^{\text{SK}} \sim \left(\dfrac{d_c}{Rm}\right)^{1/2} \sim 
\left(\dfrac{\pi\,\rins}{\ell\,Rm}\right)^{1/2}\,.
\end{equation}
The skin effect (\ref{eq:skin}) thus increases with $\ell$. 

\item Beyond $\routs$, the electrical conductivity decreases exponentially and 
the local dynamo effect is rather inefficient. The magnetic field is 
dominated by the field produced in the deeper dynamo region and approaches a 
potential field \citep[e.g.][]{Wicht19}. 
The characteristic radial dependence of a potential field in the outer 
convective region predicts:
\begin{equation}
 \dfrac{E^{P,\ell}_{M}(r_o)}{E^{P,\ell}_M(\routs)} \simeq 
\left(\dfrac{\routs}{r_o}\right)^{2\ell+4}\,.
\label{eq:vacuum}
\end{equation}
\end{enumerate}

The attenuation factors (\ref{eq:skin}) and (\ref{eq:vacuum}) 
should provide idealised upper bounds of the poloidal magnetic 
energy damping since (\textit{i}) the convective flows can penetrate into the 
SSL and
(\textit{ii}) the electrical conductivity  beyond $\routs$ still allows for 
some local dynamo action. This local action is responsible for the rise in 
magnetic energy around $\routs$ at intermediate to small scales corresponding 
to $\ell>40$ (Fig.~\ref{fig:spec_l_r}\textit{c}-\textit{d}).

\begin{figure}
 \centering
 \includegraphics[width=8.3cm]{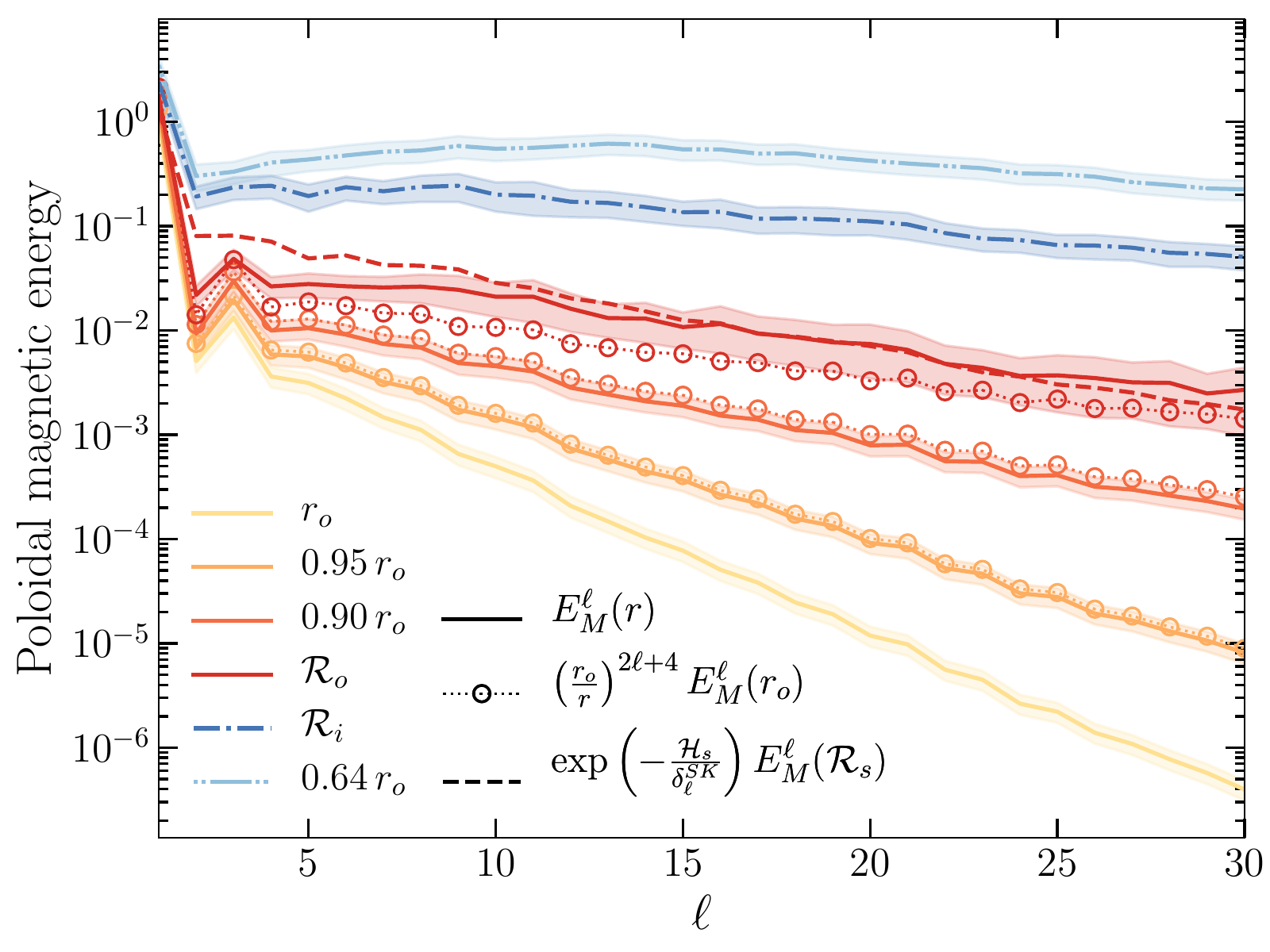}
 \caption{Time-averaged poloidal magnetic energy spectra at different 
depths up to $\ell=30$. The solid (dash-dotted)
lines correspond to the poloidal magnetic spectra above (below) 
the SSL, the circles to the downward 
continuation of the surface field (Eq.~\ref{eq:vacuum}) and the dashed line to 
the field at $\routs$ upward-continued from $\rins$ using the 
skin-depth approximation 
(Eq.~\ref{eq:skin}). The shaded regions correspond to one standard deviation 
accross the time-averaged values.}
\label{fig:specs}
\end{figure}

\begin{figure}
\centering 
 \includegraphics[width=8.3cm]{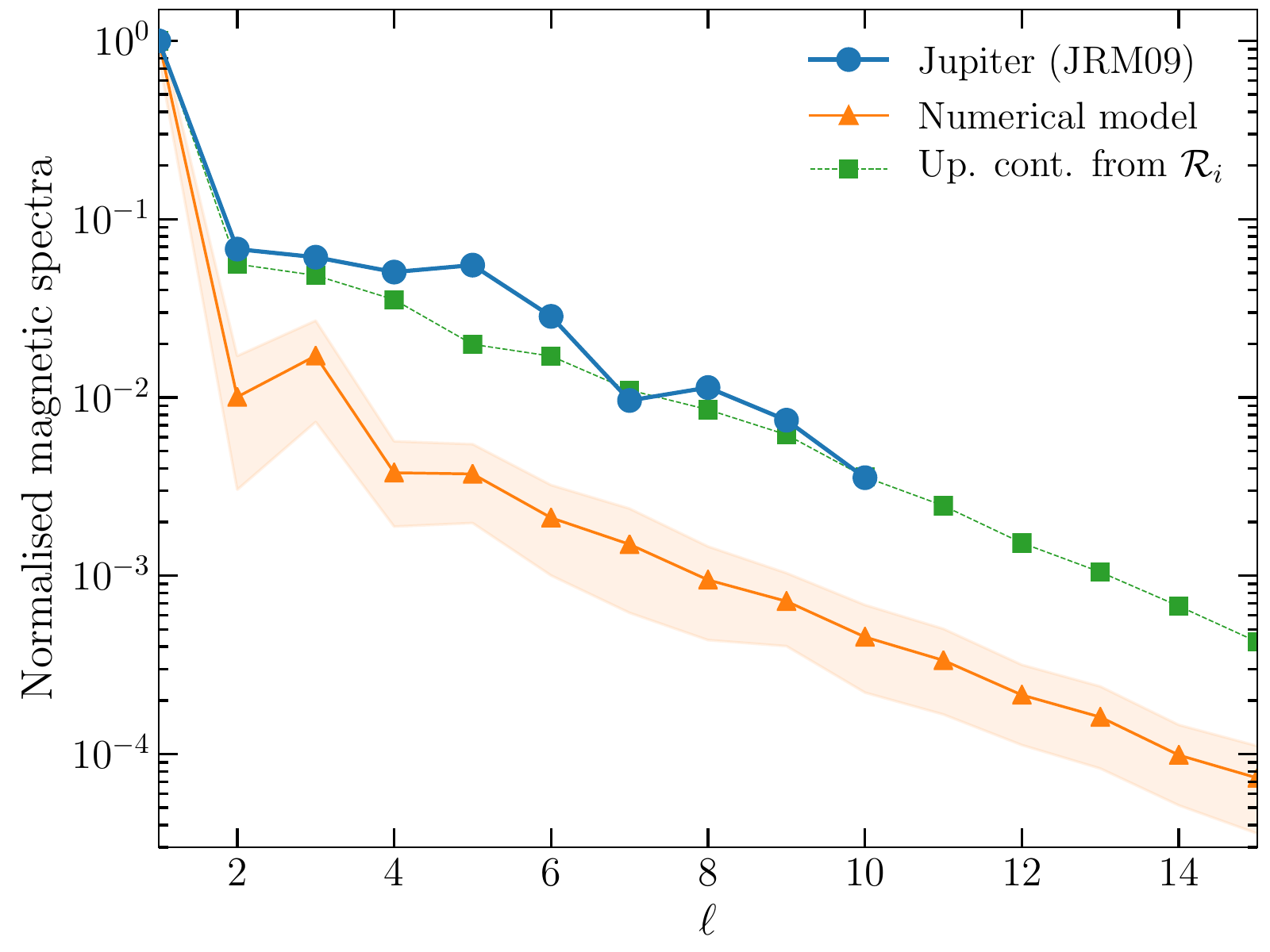}
 \caption{Normalised time-averaged magnetic spectra at the 
surface of the numerical model as well as the potential field upward 
continuation of the poloidal field at $\rins$
along with the Jovian magnetic field model JRM09 
by \cite{Connerney18} for the first $15$ harmonic degrees. The shaded area 
corresponds to one standard deviation across the time-averaged values.}
 \label{fig:compJuno}
\end{figure}

Figure~\ref{fig:specs} compares spectra of the poloidal
magnetic energy at different depths (solid lines) 
to the predictions coming from Eq.~(\ref{eq:skin}) (dashed lines) and 
Eq.~(\ref{eq:vacuum}) (circles).
Beyond $r=0.9\,r_o$, dynamo action is negligible and the poloidal
energy 
spectra closely follows the downward continuation of the surface field. 
At the top of the stable layer ($\routs=0.88\,r_o$), however, the energy 
of the downward-continued field is noticeably  smaller than the actual poloidal 
energy. The reason is the dynamo action just above or in the top part of the 
stable layer, which is also apparent in Fig.~\ref{fig:spec_l_r}\textit{c} and 
\textit{d}. Here the zonal winds induce toroidal field which is then 
converted to poloidal field by the non-axisymmetric flow components
\citep{Wicht19,Tsang20}. 

Using the poloidal field spectrum at $\rins$ combined
with the attenuation factor from the skin effect (\ref{eq:skin}) captures the 
magnetic energy spectrum at $\routs$ reasonably well.
Large scale contributions ($\ell < 15$) are overestimated, while smaller
scale contributions are slightly underestimated. The latter could be 
explained by the local dynamo action around $\routs$, which intensifies 
the field and counteracts the skin effect. 
The weaker large-scale field, on the other hand, indicates that the 
locally-induced field opposes the field produced below
the stable layer. Dipole and octupole are less affected and therefore stick out 
above the stable layer. Another reason for the discrepancy could be that 
approximating the SSL by an electrically-conducting stagnant layer is 
too simplistic despite the large degree 
of stratification considered here ($N_m/\Omega \simeq 10$).
In the deep 
interior ($r\leq \rins$), the octupole is in line with other spherical 
harmonics and the scales around $\ell\simeq 10$ nearly reach half the amplitude 
of the dipole contributions (see also 
Fig.~\ref{fig:spec_l_r}\textit{c}-\textit{d}).

\subsection{Comparison with JRM09}

\begin{figure}
 \centering
  \includegraphics[width=8.3cm]{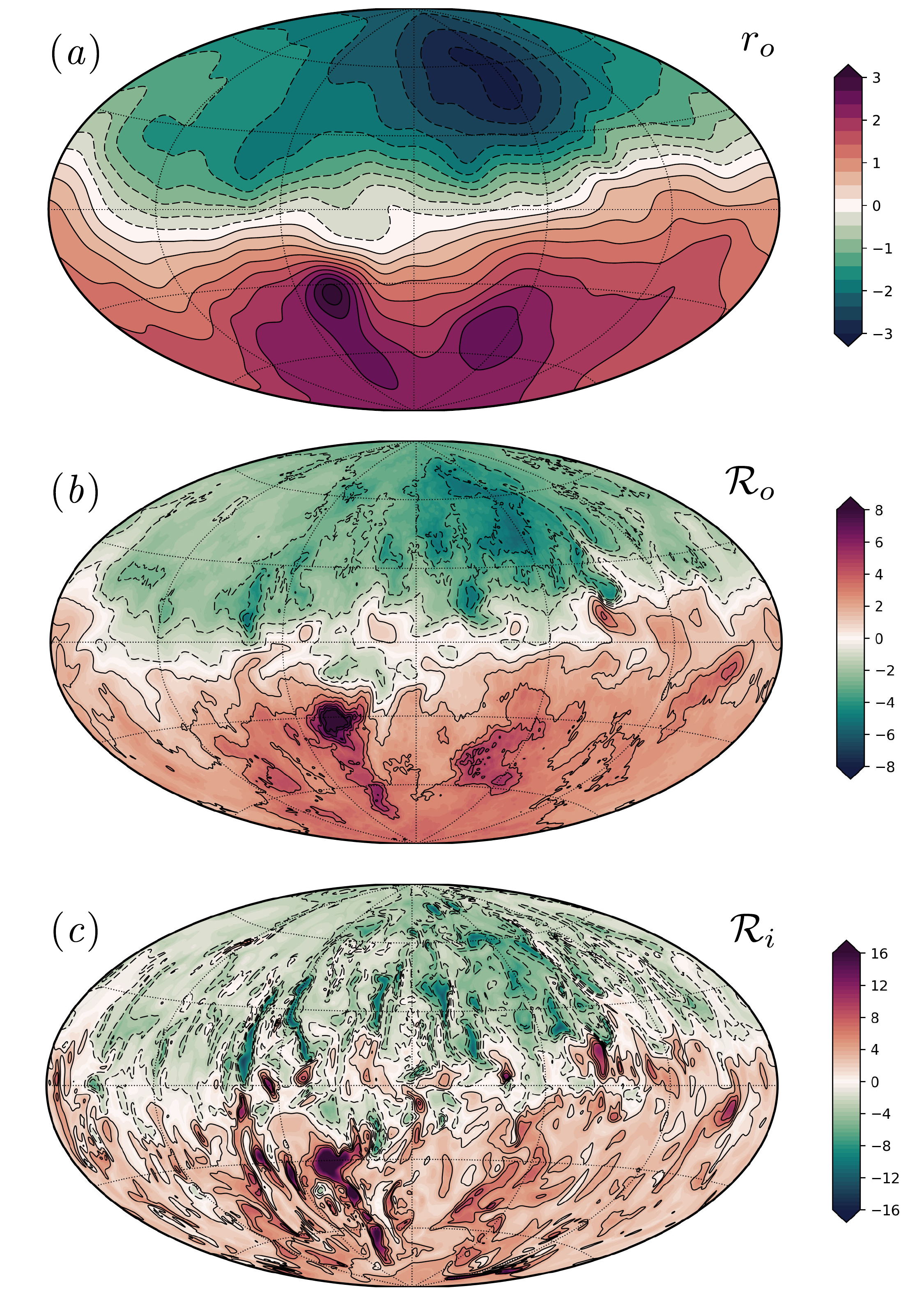}
 \caption{Hammer projection of the radial component of the magnetic field at 
 the surface (\textit{a}),  at the upper edge of the SSL $r=\routs$ 
(\textit{b}) and at the lower edge of the SSL $r=\rins$ (\textit{c}).}
 \label{fig:br}
\end{figure}

Figure~\ref{fig:compJuno} compares the normalised surface magnetic 
spectra in our simulation with the Jovian magnetic field model JRM09 by 
\cite{Connerney18}. The relative energy contained in the non-dipolar components 
is roughly one order of magnitude lower in the simulation than in the JRM09 
model. The surface 
magnetic field produced by our dynamo model is thus too dipolar, as is 
illustrated by Fig.~\ref{fig:br}, which shows the radial component of the 
magnetic field at different depths for a snapshot of our simulation. The strong 
difference between northern and southern field in JRM09 (see 
Fig.~\ref{fig:BrJup}) is not present in the simulation. There are some strong 
localised flux patches in our model, but they are 
more evenly distributed and do not stand out as clearly as in JRM09. From the 
many small scale patches at the bottom of the SSL (panel \textit{c}), the 
strongest can still be identified at the top of the SSL (panel \textit{b}) and 
are the origin of the larger scale patches at the outer boundary (panel 
\textit{a}).

Figure~\ref{fig:compJuno} also shows an upward continuation of the 
poloidal magnetic field at the base of the stable layer 
$\rins$ using Eq.~(\ref{eq:vacuum}).
This potential field approximation provides a theoretical estimate for the 
end-member attenuation when skin effect and dynamo action above the stable layer 
would be weak. The decent similarity of this approximation to the JRM09
spectrum could indicate that both are too strong in our simulation.


\subsection{Force balances}

\begin{figure*}
 \centering
 \includegraphics[width=0.95\textwidth]{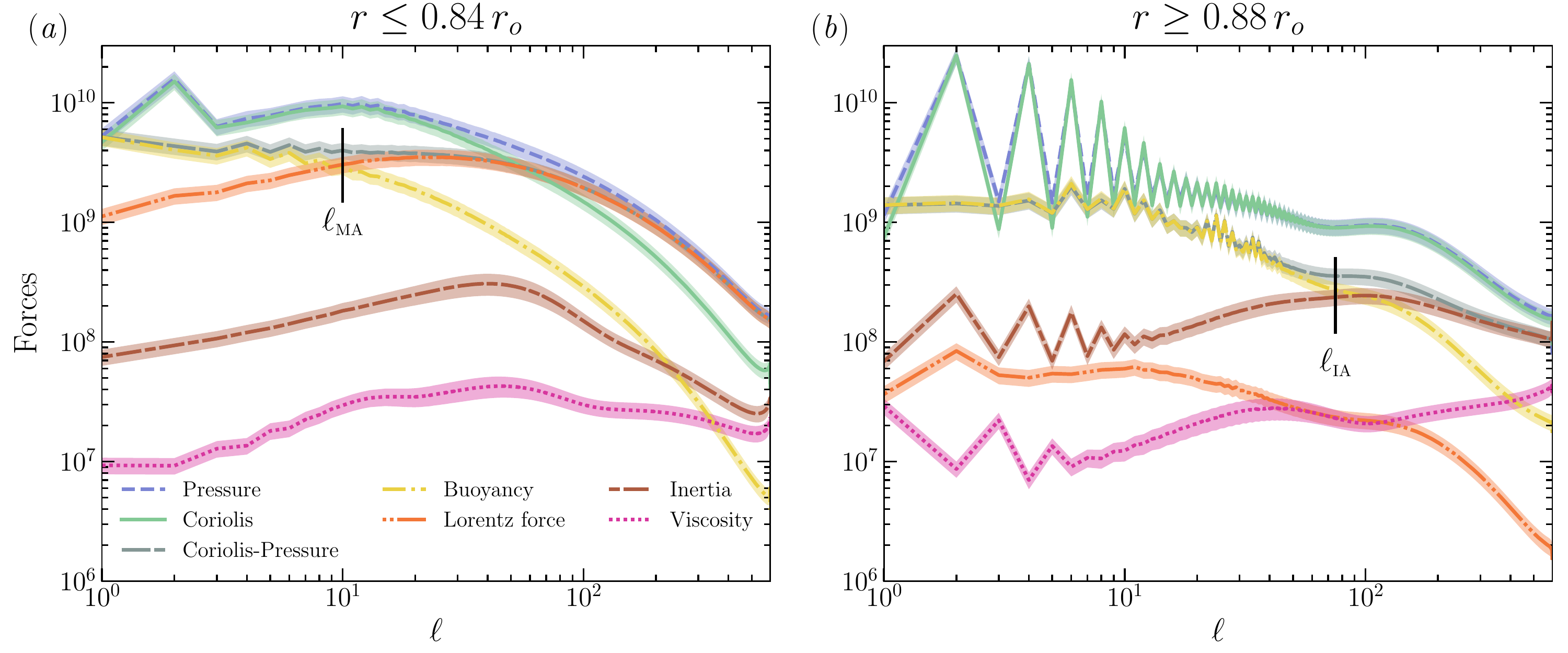}
 \caption{Time-averaged force balance spectra as a function of the harmonic 
degree integrated over the metallic core (\textit{a}) and over the 
molecular envelope (\textit{b}). The shaded area correspond 
to one standard-deviation across the mean. The vertical segments mark the 
location of the so-called ``cross-over lengthscales'' where three forces are 
in balance \citep[see][]{Aubert17,Schwaiger21}.}
 \label{fig:forces}
\end{figure*}

We now turn to examining the forces that govern the numerical dynamo model. To 
do so, we resort to the analysis of the spectral decomposition of forces 
introduced by \cite{Aubert17} and \cite{Schwaiger19}. Each force vector 
$\vec{f}$ is expanded in vector spherical harmonics
\begin{equation}
 \vec{f}(r,\theta,\phi,t) = \sum_{\ell=0}^{\ell_{\text{max}}} 
\mathcal{Q}_\ell^m 
Y_\ell^m 
\vec{e_r} + \mathcal{S}_\ell^m \,r \vec{\nabla} Y_\ell^m + \mathcal{T}_\ell^m 
\vec{r}\times \vec{\nabla} Y_\ell^m,
\end{equation}
where $\vec{r}$ is the vector along the radial direction and 
$Y_\ell^m(\theta,\phi)$ is the spherical harmonic of degree $\ell$ and order 
$m$. The energy of the vector $\vec{f}$ is then retrieved by the following 
identity
\[
\begin{aligned}
 F^2 & = \int_{V} \vec{f}^2 \mathrm{d}V, \\
 & = 2 \int_{r_i}^{r_o} \sum_{\ell=0}^{\ell_\text{max}} 
\sideset{}{'}\sum_{m=0}^{\ell} |\mathcal{Q}_\ell^m|^2 + 
\ell(\ell+1)\left(|\mathcal{S}_\ell^m|^2+|\mathcal{T}_\ell^m|^2\right)\,
r^2 \mathrm {d } r\,,
\end{aligned}
\]
where the prime on the summation over the order $m$ indicates that the $m=0$ 
coefficient is multiplied by one half. To examine the spectral 
distribution of the forces, the above expression is rearranged as 
follows:
\begin{equation}
 F^2 = \sum_\ell \mathcal{F}_\ell^2(r_i,r_o),
\end{equation}
where
\begin{equation}
 \mathcal{F}_\ell^2(r_b,r_t) =  2 \int_{r_b}^{r_t}  
\sideset{}{'}\sum_{m=0}^\ell 
|\mathcal{Q}_\ell^m|^2 
+ 
\ell(\ell+1)\left(|\mathcal{S}_\ell^m|^2+|\mathcal{T}_\ell^m|^2\right)\, r^2 
\mathrm{d}r\,.
\end{equation}
We adapt the bounds of the radial integration $r_b$ and $r_t$ to 
either focus on the metallic core or on the convective envelope. 
Figure~\ref{fig:forces} shows the time-averaged force balance spectra 
$\overline{F_\ell}(r_i,\rins)$ (left) and $\overline{F_\ell}(\routs,r_o)$ 
(right).
We find a primary geostrophic balance (QG) between pressure gradient and 
Coriolis force at large scales with $\ell<70$. At smaller scales, the pressure 
gradient is superseded  by Lorentz forces in a magnetostrophic balance 
(MS) \citep{Aurnou17}. Beyond this 
primary balance, the difference between pressure gradient and Coriolis force, 
termed
\emph{ageostrophic Coriolis force}, is in balance with buoyancy at large scales 
($\ell < 10$) and with Lorentz force at small scales. Inertia and viscosity are 
respectively one and two orders of magnitude below this first-order
balance. This forms the so-called \emph{QG-MAC 
balance} (Magneto, Archimedean, Coriolis) introduced theoretically by 
\cite{Davidson13} and identified in reduced numerical models by \cite{Calkins18}
and full dynamo simulations by \cite{Schwaiger19}. 
This hierarchy of forces is structurally similar to the ones obtained in the 
geodynamo models of 
\cite{Schwaiger19} when the magnetic energy exceeds the kinetic one. We note 
that the separation between Lorentz force and inertia is of comparable amplitude 
to the ratio of magnetic and kinetic energies (see 
Fig.~\ref{fig:radprofs}\textit{a}). 

In the molecular envelope, the leading-order quasi-geostrophic equilibrium is 
accompanied by a secondary balance between ageostrophic Coriolis force and 
buoyancy up to $\ell \simeq 70$ and between ageostrophic Coriolis force and 
inertia beyond. Because of the decrease of electrical conductivity, Lorentz 
forces play a much weaker role and have a comparable amplitude to 
the viscous force. The convective flows in the outer convective layer therefore  
obey the 
so-called \emph{QG-IAC balance} (Inertia, Archimedean, Coriolis) derived by 
\cite{Cardin94} in the context of quasi-geostrophic convection \cite[see 
also][]{Aubert01,Gillet06,Gastine16}.

The spectral representations shown in Fig.~\ref{fig:forces} also reveal 
the cross-over lengthscales \citep{Aubert17,Schwaiger21} 
defined by the 
harmonic degree at which at least two forces are of equal amplitude.
Of particular interest are the intersections between buoyancy and Lorentz 
forces in the metallic core and between buoyancy and inertia in the molecular 
envelope. The respective degrees $\ell_{\text{MA}}\approx 10$ and 
$\ell_{\text{IA}} \approx
75$, marked by the two vertical segments in Fig.~\ref{fig:forces}, 
characterise the lengthscale of optimal QG-MAC and QG-IAC balances. As already 
reported by \cite{Aubert17}, those cross-over 
lengthscales are in good agreement with 
the dominant lengthscale of convection $\hat{\ell}$ defined by the peak of 
the non-axisymmetric kinetic energy (Fig.~\ref{fig:spec_l_r}\textit{a}).

%

This implies that the most energetic convective features are controlled by a 
QG-MAC balance in the metallic core and a QG-IAC balance in the external 
convective region, two force balance hierarchies expected to hold in the 
interiors of gas giants.

\subsection{Zonal and meridional flows}

\begin{figure}
 \centering
 \includegraphics[width=8.3cm]{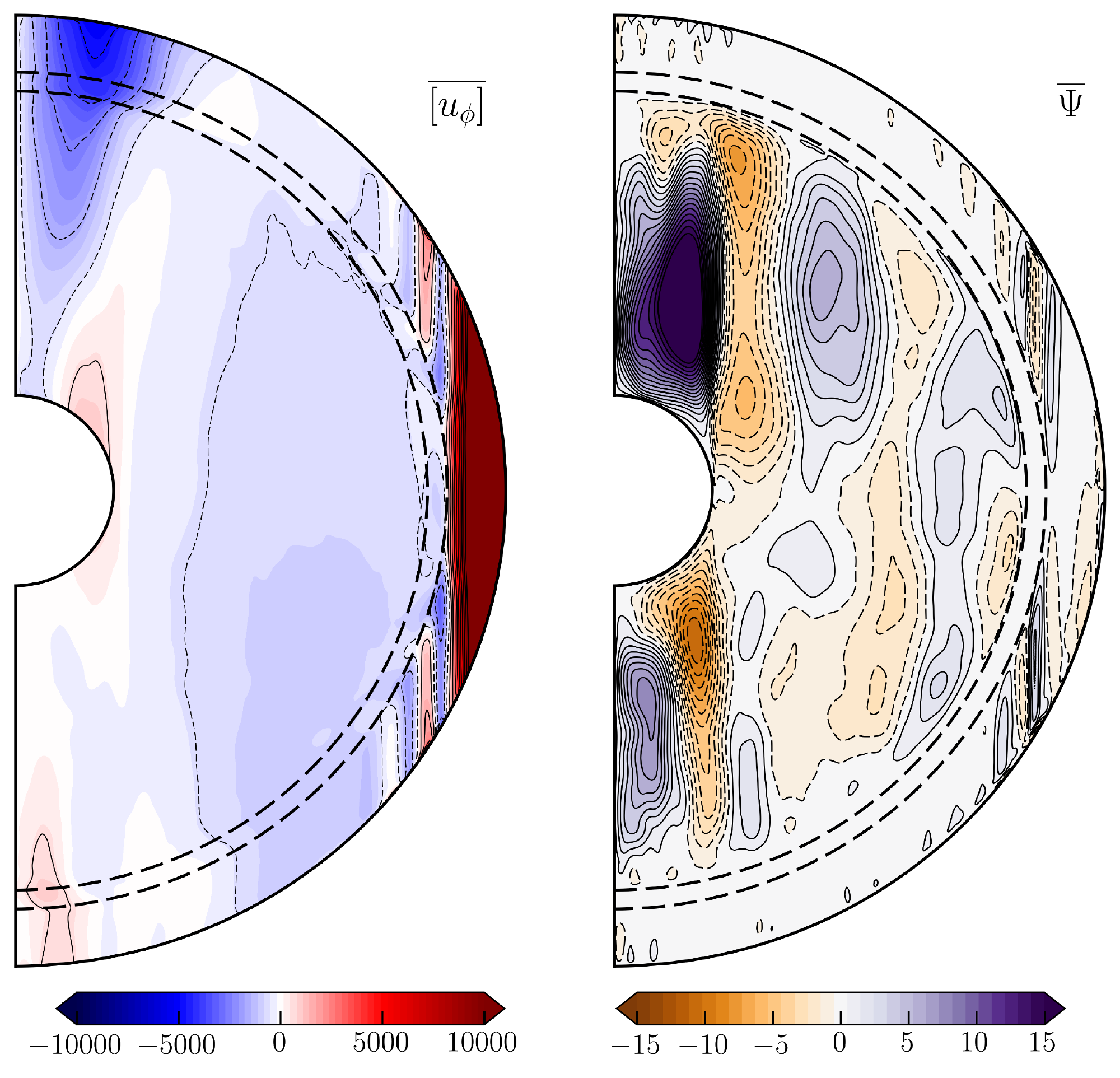}
 \caption{(\textit{a}) Time-averaged zonal flows $\overline{[u_\phi]}$. 
(\textit{b}) Time-averaged stream function of the 
meridional circulation $\overline{\Psi}$. Solid (dashed) contour lines 
correspond to clockwise (counter clockwise) meridional circulation. In both 
panels, the dashed half circles mark the bounds of the SSL $\rins$ and 
$\routs$.}
\label{fig:vp_psi}
\end{figure}

We now examine the structure of the axisymmetric flows produced in this 
numerical dynamo model. Figure~\ref{fig:vp_psi} shows the time-averaged zonal 
flow $\overline{[u_\phi]}$ and the stream function $\overline{\Psi}$  
associated with the meridional circulation defined by

\[
 \rbar\,\vec{u_m} = \vec{\nabla} \times (\rbar \,\Psi \vec{e_\phi})\,,
\]
where $\vec{u_m}=([u_r],[u_\theta])$ is the meridional circulation vector and 
$\vec{e_\phi}$ is the unit vector in the $\phi$ direction.
In the molecular envelope, the zonal motions are dominated by a strong 
prograde equatorial jet.
On each side of the equatorial jet we find two retrograde and two prograde 
secondary jets. The innermost prograde jets, located at about $40^\circ$ 
latitude 
north and south, are particularly faint. 
This jet system persisted over our simulation time, which is equivalent to 
about $8400$ rotations. 
Zonal winds at high latitudes form broader structures, which are often 
dominated by 
thermal wind features and change over time. The deeper convective region 
exhibits much weaker differential rotation \citep[see][]{Jones14}.

The meridional flows are one to two orders of magnitude weaker than the 
typical non-axisymmetric convective flows. In the external convective region, 
it forms pairs of equatorially 
anti-symmetric cells elongated along the rotation axis. 
The cells are highly correlated with the zonal jets.
The stable stratification effectively prevents the meridional circulations from 
penetrating the SSL. 

Within the metallic interior, the meridional circulation resides on more 
intricate columnar cellular patterns which also show some correlation with the 
zonal winds.

\begin{figure*}
 \centering
    \includegraphics[width=0.99\textwidth]{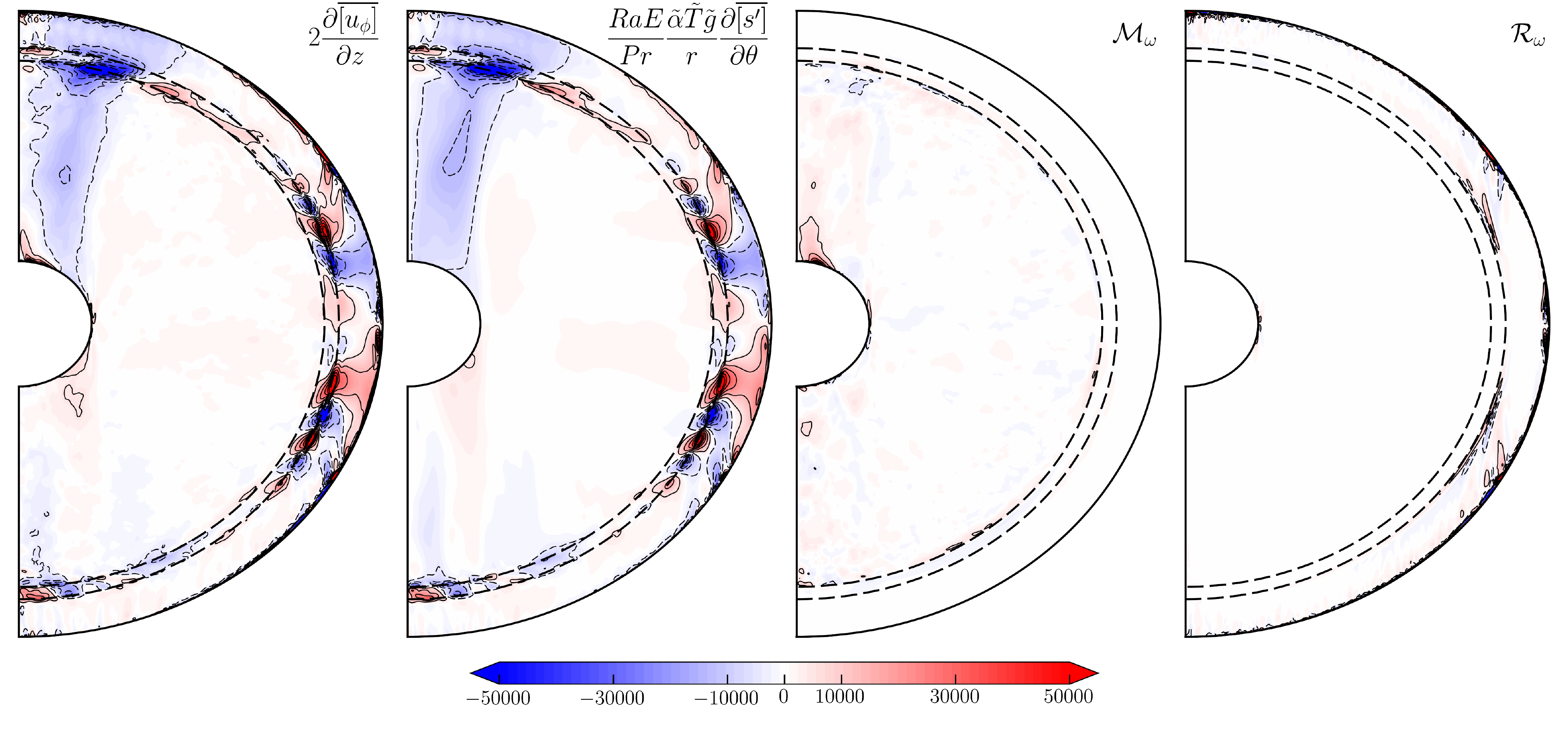}
   \caption{Meridional cuts of the time-averaged terms that enter the thermal 
wind balance (\ref{eq:thwind}). Because of its much weaker amplitude, the 
viscous contribution $\mathcal{V}_\omega$ entering Eq.~(\ref{eq:thwind}) has 
been omitted. The dashed half circles mark the bounds of the SSL $\rins$ and 
$\routs$.}
\label{fig:thWind}
  \includegraphics[width=0.99\textwidth]{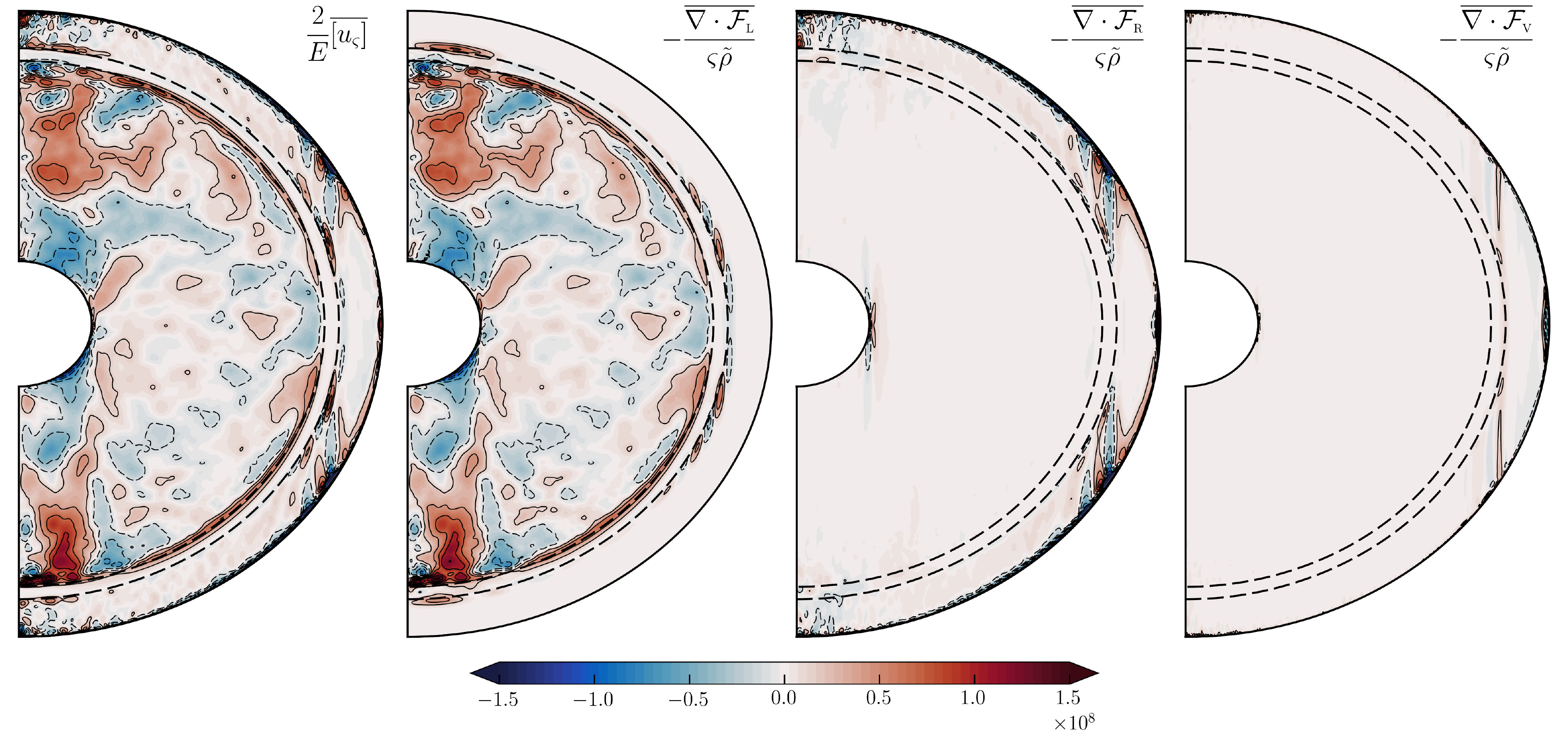}
 \caption{Meridional cuts of the time-averaged terms that enter the angular 
momentum transport equation (\ref{eq:zon}). The dashed half circles mark the 
bounds of the SSL $\rins$ and $\routs$.}
\label{fig:vp_bal}
\end{figure*}

In order to understand the quenching of the jets and the correlation with the 
meridional circulation, we consider two fundamental equations. The first one is 
the thermal wind equation, which can be derived from the azimuthal component of 
the curl of the Navier-Stokes equation (\ref{eq:NS}):

\begin{equation}
\begin{aligned}
 \dfrac{D \omega_\phi}{D t} 
 =& \dfrac{2}{E}\dfrac{\partial 
u_\phi}{\partial z} -\dfrac{Ra}{Pr}\dfrac{\abar\tbar \gbar}{r}\dfrac{\partial 
s'}{\partial\theta}+\cyl\vec{\omega}\cdot\vec{\nabla}\left(\dfrac{
u_\phi } { \cyl } \right)-\omega_\phi\vec{\nabla}\cdot\vec{u}\\
&+\vec{e_\phi}\cdot\vec{\nabla}\times 
\left(\dfrac{\vec{j}\times\vec{B}}{E Pm\,\rbar}\right)+
\vec{e_\phi}\cdot\vec{\nabla}\times\left(\dfrac{\vec{\nabla}\cdot\tens{S}}{\rbar
} \right)\,.
\end{aligned}
\label{eq:vortzon}
\end{equation}
Here $\omega_\phi=\vec{e_\phi}\cdot\vec{\nabla}\times\vec{u}$ and 
$\cyl=r\sin\theta$ denotes the cylindrical radius. When averaging over time and 
azimuth, Eq.~(\ref{eq:vortzon}) yields

\begin{equation}
2 \dfrac{\partial \overline{[u_\phi]}}{\partial z} = \dfrac{Ra 
E}{Pr}\dfrac{\abar \tbar \gbar}{r}\dfrac{\partial \overline{[s']}}{\partial 
\theta} + \mathcal{R}_\omega + \mathcal{M}_\omega +\mathcal{V}_\omega\,.
\label{eq:thwind}
\end{equation}
In the above equation, $\mathcal{R}_\omega$ is a nonlinear term defined by

\[
 \mathcal{R}_\omega = E\left(
\overline{\left[\vec{u}\cdot\vec{\nabla}\omega_\phi\right]} 
-\cyl\overline{\left[\vec{\omega}\cdot\vec{\nabla}\dfrac{u_\phi}{\cyl}\right]} 
-\dfrac{\mathrm{d}\ln\rbar}{\mathrm{d} r}\overline{[u_r\omega_\phi]}\right),
\]
where the three contributions entering the right-hand-side respectively 
correspond to 
advection, stretching and compressional sources of vorticity. 
$\mathcal{M}_\omega$ and $\mathcal{V}_\omega$ denote the magnetic and viscous 
stresses defined by

\[
 \mathcal{M}_\omega =-\dfrac{1}{Pm}\overline{\left[\vec{e_\phi}\cdot
 \vec{\nabla}\times\left(\dfrac{\vec{j}\times\vec{B}}{\rbar}\right)\right]}\,,
\]
and
\[
 \mathcal{V}_\omega = -E \overline{\left[\vec{e_\phi} 
\cdot 
\vec{\nabla}\times\left(\dfrac{\vec{\nabla}\cdot\tens{S}}{\rbar}\right)\right]}
\,.
\]
Figure~\ref{fig:thWind} shows meridional cuts of the different terms in 
Eq.~(\ref{eq:thwind}). The axial gradient of $\overline{[u_\phi]}$ 
almost perfectly balances the latitudinal gradient of entropy, with small 
remaining contributions of magnetic winds $\mathcal{M}_\omega$ inside the 
tangent cylinder and from 
inertia close to the upper edge of the SSL around $45^\circ$ latitude. The 
classical thermal wind balance

\begin{equation}
 2 \dfrac{\partial \overline{[u_\phi]}}{\partial z} \approx \dfrac{Ra 
E}{Pr}\dfrac{\abar \tbar \gbar}{r}\dfrac{\partial \overline{[s']}}{\partial 
\theta}\,,
\label{eq:thWindShort}
\end{equation}
is hence realised to a high degree of fidelity, indicating that
Lorentz forces have no direct impact on the $z$-variations of the zonal flows.
The strongest latitudinal entropy gradients are found at the upper edge of 
the SSL between $20^\circ$ and $45^\circ$ latitude, where the alternating 
zonal flows rapidly decay. The entropy gradients are much weaker in the middle 
of the external convective region where the zonal winds remain nearly 
geostrophic. The braking of $\overline{[u_\phi]}$ at 
$\routs$ is accommodated by intense localised entropy variations.

\begin{figure*}
 \centering
 \includegraphics[width=0.99\textwidth]{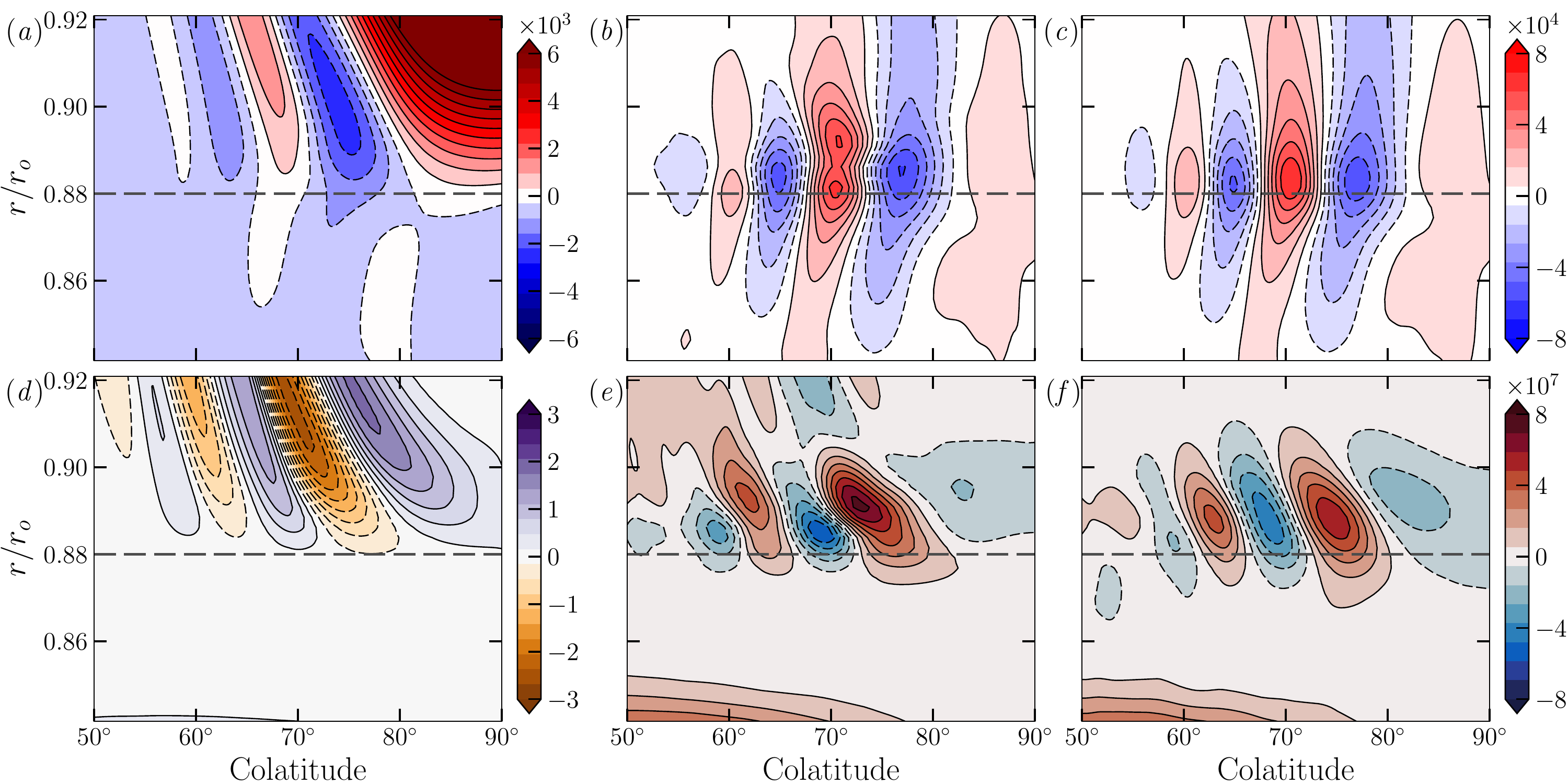}
 \caption{Zoomed-in insets of Fig.~\ref{fig:vp_psi}, \ref{fig:thWind} and 
\ref{fig:vp_bal} for $r\in[\rins,0.92\,r_o]$ and 
$\theta\in[50^\circ,90^\circ]$. (\textit{a}) Time-averaged zonal flows 
$\overline{[u_\phi]}$. (\textit{b}) Time-averaged axial gradient of the zonal 
flows $2\,\partial \overline{[u_\phi]}/\partial z$. (\textit{c}) Time-averaged 
meridional gradient of temperature $(Ra E/Pr)(\abar\tbar\gbar/r)\partial 
\,\overline{[s']}/\partial \theta$. (\textit{d}) Time-averaged stream function 
of the meridional circulation $\overline{\Psi}$. (\textit{e}) Time-averaged 
axisymmetric component of Coriolis force  $2\,\overline{[u_\cyl]}/E$.
(\textit{f}) Time-averaged axisymmetric $\phi$-component of the Lorentz force 
$-\overline{\vec{\nabla}\cdot\mathcal{F}_L}/\rbar\cyl$. (\textit{b}) and 
(\textit{c}) correspond to the dominant terms of thermal wind balance 
(\ref{eq:thWindShort}) shown in Fig.~\ref{fig:thWind}. (\textit{e}) and 
(\textit{f}) correspond to the dominant terms of the angular momentum transport 
equation (\ref{eq:zon}) shown in Fig.~\ref{fig:vp_bal}. In each panel, the 
horizontal dashed line corresponds to $r=\routs$.}
 \label{fig:blow_up}
\end{figure*}

To examine the force balance that sustains the meridional circulation 
pattern, we now consider the zonal component of the Navier-Stokes equation 
(\ref{eq:NS}):
\begin{equation}
  \rbar\dfrac{\partial [u_\phi]}{\partial t}+\dfrac{2}{E}\rbar[u_\cyl] = 
-\dfrac{1}{\cyl} \vec{\nabla}\cdot \vec{\mathcal{F}},
\label{eq:zon}
\end{equation}
where $u_\cyl$ corresponds to the cylindrically-radial component of the 
velocity. The angular momentum flux $\vec{\mathcal{F}}$ can be decomposed 
into three contributions,
\[
 \vec{\mathcal{F}} =  \vec{\mathcal{F}}_\text{R} +  
\vec{\mathcal{F}}_\text{M} + \vec{\mathcal{F}}_\text{V},
\]
accounting for Reynolds, Maxwell and viscous stresses
\[
 \vec{\mathcal{F}}_\text{R}=\rbar \cyl [\vec{u}u_\phi],\ 
 \vec{\mathcal{F}}_\text{M}=-\dfrac{\cyl [\vec{B}B_\phi]}{EPm},\ 
  \vec{\mathcal{F}}_\text{V}=-\rbar 
\cyl^2\vec{\nabla} \left(\dfrac{[u_\phi]}{\cyl}\right)\,.
\]
On time-average, 
the flow perpendicular to the rotation axis $\overline{[u_\cyl]}$ responds to 
the imbalance between those 
different axial forces, a physical phenomenon termed ``geostrophic pumping'' 
by \cite{McIntyre98}. Figure~\ref{fig:vp_bal} shows meridional 
cuts of the different time-averaged contributions to Eq.~(\ref{eq:zon}).
In the metallic interior, the axisymmetric components of the Lorentz and 
Coriolis 
forces balance each other almost perfectly with secondary contributions of 
inertia close to the inner boundary. This pattern is typical of 
rapidly-rotating convection when Lorentz forces play a dominant role in 
the force balance \citep[see, e.g.][his Fig.~7]{Aubert05}. 

The situation in the external convective region (beyond $\routs$) is more 
intricate. The Reynolds stresses that maintain the observed alternating zonal 
jet pattern mainly act in the upper parts of the external convective layer, 
where the typical convective flows are more vigorous. This driving is 
compensated partly by viscous stresses in the intense shear regions and 
partly by Maxwell stresses at the bottom of the external convective region 
($r\gtrsim \routs$) where the electrical conductivity is still sizeable.
Maxwell stresses play a negligible role for the equatorial jet since 
it penetrates less deep. However, they are definitely important for braking the 
flanking jets.

\begin{figure}
 \centering
 \includegraphics[width=0.48\textwidth]{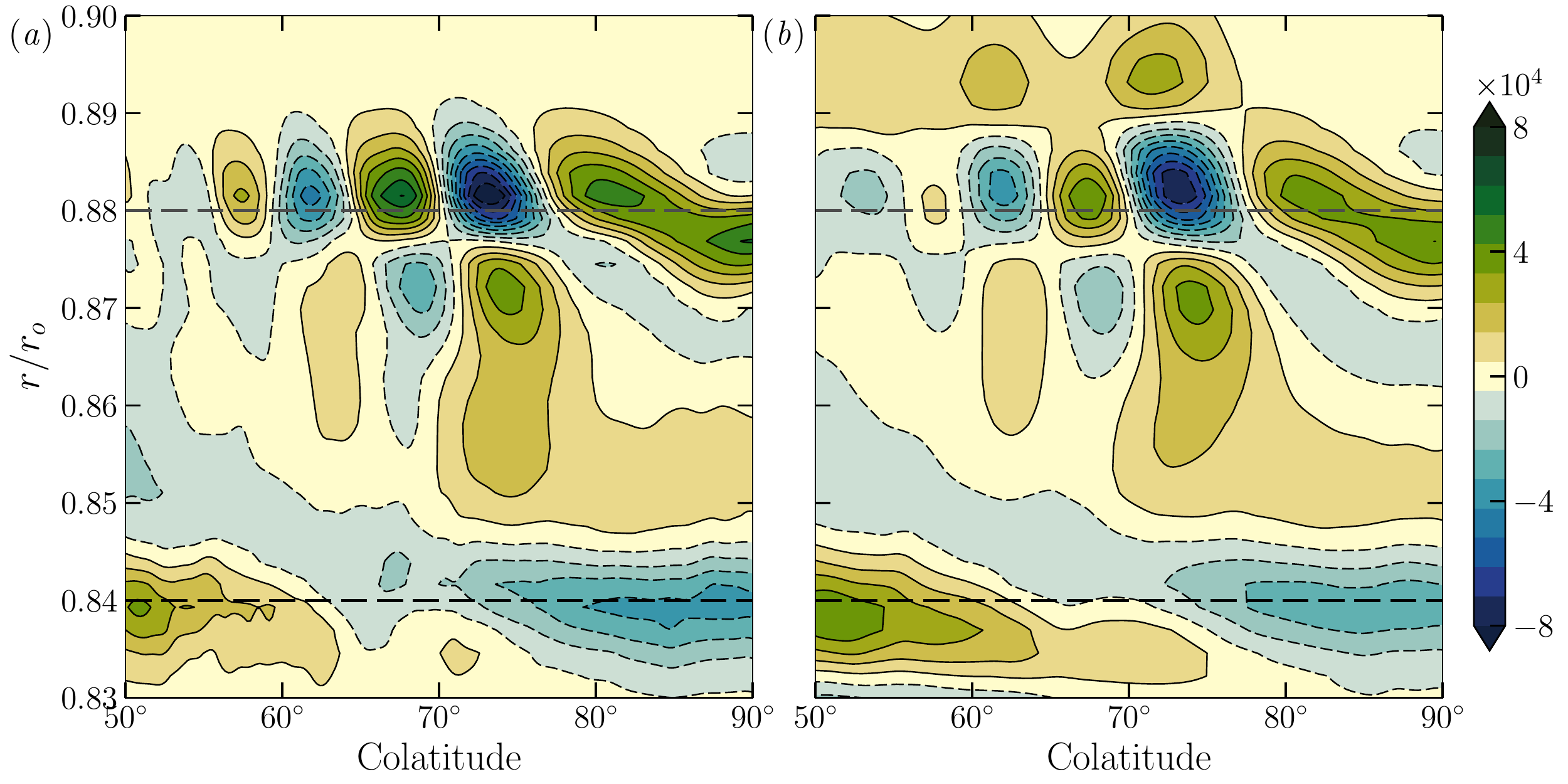}
 \caption{Zoomed-in insets for $r\in[0.83\,r_o,0.9,r_o]$ and 
$\theta\in[50^\circ,90^\circ]$. (\textit{a}) Time-averaged axisymmetric 
advection of the entropy background by the meridional flow $\rbar \tbar\, 
\overline{[u_r]} \mathrm{d} \sbar /\mathrm{d} r$. (\textit{b}) Time-averaged 
radial part of the entropy diffusion $1/Pr\,\vec{\nabla}\cdot(\rbar 
\tbar \vec{\nabla} \overline{[s']} \cdot\vec{e_r})$. In both panels the 
horizontal dashed lines correspond to $r=\rins$ and $r=\routs$.}
\label{fig:radialHeat}
\end{figure}

At the upper edge of the SSL, the delicate balance between Maxwell and Reynolds 
stresses drives a meridional circulation pattern which slightly penetrates the 
stable layer. This is the main player in establishing the latitudinal entropy 
variation that explains the quenching of the zonal winds. 
Figure~\ref{fig:blow_up} illustrates the interesting dynamics in the region 
where the jets touch the upper edge of the stable layer.  
The upper row highlights the importance of the thermal wind balance 
(Eq.~\ref{eq:thWindShort}) for limiting the depth of the flanking jets. The 
$z$-variation (panel \textit{b}) in the zonal flows (panel \textit{a}) are 
nearly perfectly explained by the thermal wind term that depends on  
axisymmetric latitudinal entropy variations (panel \textit{c}). 

The lower row of Fig.~\ref{fig:blow_up} illustrates how the stable 
stratification effectively prevents the meridional circulation (panels 
\textit{d} and \textit{e}) from penetrating the SSL. Azimuthal Lorentz force
(panel \textit{f}) shapes the meridional circulation pattern (panel \textit{e}) 
according  to Eq.~(\ref{eq:zon}). This force is a direct result of the electric 
currents induced by the zonal winds \citep{Wicht19a}.

Figure~\ref{fig:radialHeat} shows that the time-averaged advection of the 
entropy background $\mathrm{d}\sbar/\mathrm{d}r$ by the meridional flow 
$\overline{[u_r]}$ is balanced to a large degree by the radial 
diffusion $1/Pr\,\vec{\nabla}\cdot(\rbar \tbar \vec{\nabla} \overline{[s']}
\cdot\vec{e_r})$. Other entropy transport contributions are of secondary 
importance close to the SSL.
This implies that the meridional circulation cells which scratch the
upper edge of the SSL build up the local latitudinal entropy gradients
visible in panel (\textit{c}) of Fig.~\ref{fig:blow_up}.

\begin{figure}
 \centering
 \includegraphics[width=8.3cm]{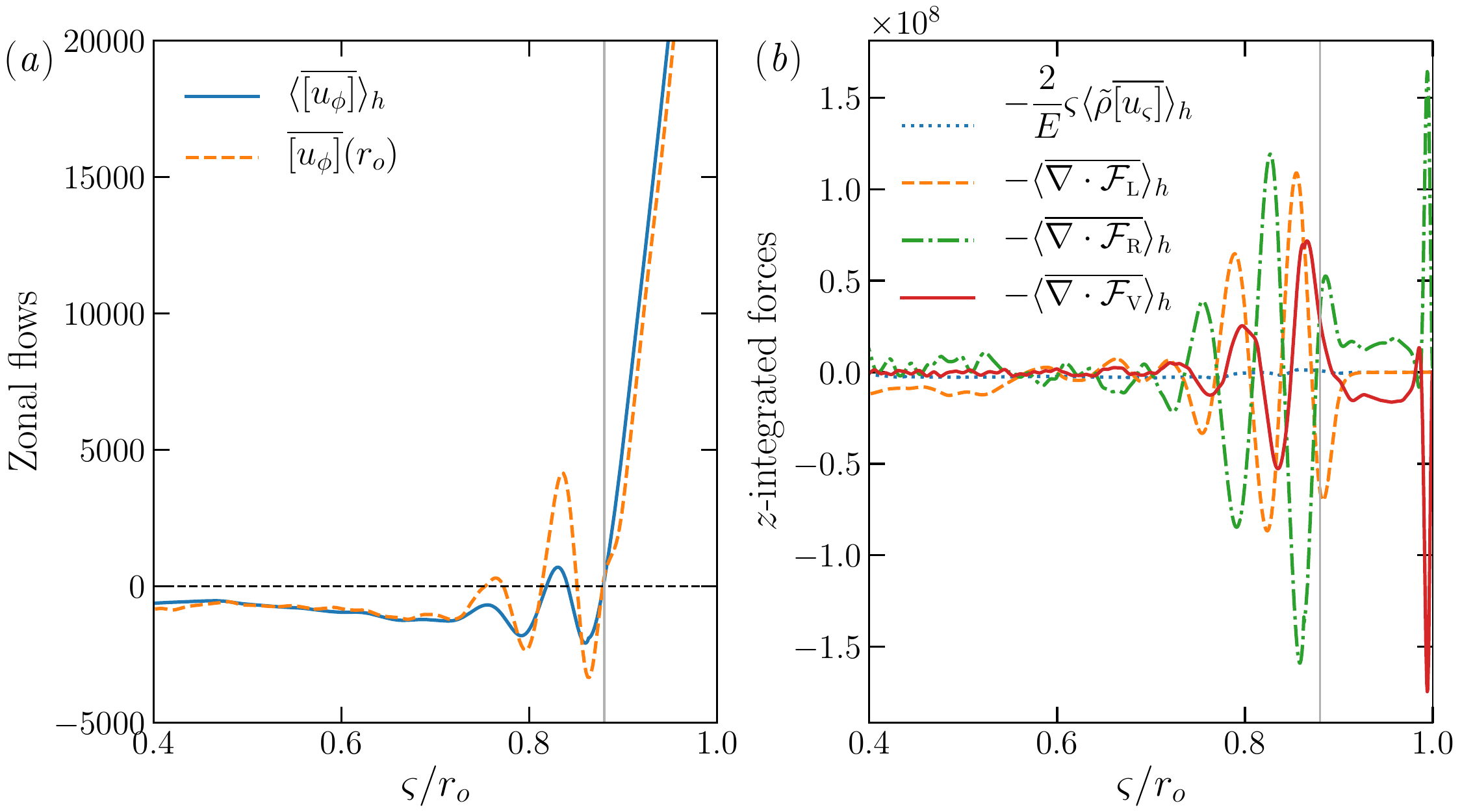}
 \caption{(\textit{a}) Time-averaged surface zonal flows in the Northern 
hemisphere (dashed lines) and 
geostrophic zonal flows (solid lines) as a function of the normalised 
cylindrical radius. (\textit{b}) Time-averaged axial torques integrated over 
cylinders as a function of $\cyl/r_o$. The vertical lines correspond to the 
upper edge of the SSL $\routs$.}
 \label{fig:forces_cyl}
\end{figure}

To study the roles played by the Lorentz force and viscosity 
in controlling the amplitude of the zonal jets, we integrate 
Eq.~(\ref{eq:zon}) over axial cylinders for the fluid regions above the 
middle of the SSL.

\begin{equation}
 \dfrac{2}{E}s \left\langle \tilde{\rho} \overline{ [u_s]} \right\rangle_h =
-\left\langle\overline{\vec{\nabla}\cdot\vec{\mathcal{F}_\text{R}}}
\right\rangle_h
-\left\langle\overline{\vec{\nabla}\cdot\vec{\mathcal{F}_\text{M}}}
\right\rangle_h
-\left\langle\overline{\vec{\nabla}\cdot\vec{\mathcal{F}_\text{V}}}
\right\rangle_h.
\label{eq:vpgeos}
\end{equation}
The operator $\langle f \rangle_h$ is defined by

\[
 \left\langle f\right \rangle_h = \dfrac{1}{h^{+}-h^{-}}\int_{h^{-}}^{h^{+}} 
f(\cyl,z)\, \mathrm{d} z\,.
\]
where the bounds of integration $h^{+}$ and $h^{-}$ depend on the
radius $\rcs=\frac{1}{2}(\routs+\rins)$. For $\cyl \geq \rcs$,
$h^{\pm} = \pm\sqrt{r_o^2 -\cyl^2}$, while the integration bounds are 
restricted to the Northern hemisphere for $s < \rcs$, i.e. 
$h^{+}=\sqrt{r_o^2-\cyl^2}$  and $h^{-} = \sqrt{\rcs^2-\cyl^2}$.

Figure~\ref{fig:forces_cyl}\textit{a} shows the geostrophic component of the 
zonal flows, $\langle\overline{[u_\phi]}\rangle_h$, along 
with the surface profile $\overline{[u_\phi]}(r_o)$, while 
Fig.~\ref{fig:forces_cyl}\textit{b} portrays the different 
time-averaged axial torques which enter Eq.~(\ref{eq:vpgeos}).
As already observed in Fig.~\ref{fig:vp_psi}, the upper edge of the SSL marks a 
clear separation of the zonal flow morphology. For $s > \routs$, the 
geostrophic component of the zonal flows closely follows the surface profile, 
indicating the high degree of geostrophy of the main prograde equatorial jet. 
Because of the decay of the zonal flows in the SSL, the secondary jets between 
$0.7\,r_o < \cyl < \routs$ feature a much weaker geostrophic component. 
This dynamical change in the vicinity of $\routs$ is also recovered in the 
spatial distribution of the axial torques. The strong prograde equatorial jet 
is driven by positive Reynolds stresses equilibrated by viscosity, while the 
geostrophic part of the  secondary alternating jets are driven by undulating 
Reynolds stresses balanced by a combination of Lorentz and viscous torques.
The cylindrical integration of the Coriolis term vanishes indicating the 
cancellation of the mass flux over the considered fluid domain $r \geq \rcs$.

\section{Discussion and conclusion}

\label{sec:disc}

Several recent Jupiter's interior models suggest that Helium demixing could 
happen in a thin layer located close to the transition to metallic 
hydrogen \citep[e.g.][]{Militzer16,Wahl17,Debras19}.
To examine the effects of such a layer, we have developed the 
first global dynamo model of Jupiter that incorporates a stably-stratified layer 
between $0.82\,R_J$ and $0.86\,R_J$. The chosen degree of 
stratification characterised by the ratio of the Brunt-V\"ais\"al\"a frequency 
to the rotation rate is rather strong with $N_m/\Omega\simeq 10$ to ensure that 
convection would not penetrate through the stably-stratified layer (SSL).
Such an SSL effectively separates the dynamics of the regions below and above. 
Previous simulations without such a layer suggest that only the equatorial jet 
is compatible with Jupiter-like dynamo action \citep{Jones14,Gastine14a}. 
Stronger flanking jets would always penetrate into the highly-conducting 
interior and lead to too complex fields unlike Jupiter 
\citep{Duarte13,Dietrich18}. For the first time, we show that the SSL allows 
flanking jets to develop while maintaining dipole-dominated dynamo action. 
The flanking jets only extend up to $\pm 40^\circ$ degree in latitude and are 
weaker than observed on Jupiter.

The dynamics below and above the SSL obey different underlying force 
balances. By directly measuring the spectral distribution 
of forces, we have shown that the metallic region is controlled by a triple 
force balance between the non-geostrophic part of Coriolis force, buoyancy and 
Lorentz forces, with secondary contributions of inertia and viscosity. This 
forms the so-called \emph{QG-MAC} balance which has been devised by 
\cite{Davidson13}, and is expected to hold in the dynamo regions of gas giants. 
The outer convective region where the electrical conductivity drops follows 
a different force balance with dominant contributions of ageostrophic Coriolis 
force, buoyancy and inertia. This corresponds to the so-called \emph{QG-IAC} 
balance \citep[see][]{Cardin94,Aubert03,Gillet06,Gastine16}, 
a physical regime at work in convective regions of rapidly-rotating 
astrophysical bodies when the magnetic effects are negligible.
Despite diffusivities orders of magnitudes larger than in the gas giants, the 
dynamo model presented here obeys the leading order force balances expected to 
hold in Jupiter's interior.

The mechanism that prevents the jets from penetrating the SSL in our simulations 
follows the scenario outlined by \cite{Christensen20}. Where the zonal 
winds reach to high conductivities, their induction yields Lorentz forces that 
in turn drive a complex meridional circulation pattern. Where this circulation 
penetrates the SSL and encounters the strong background stratification, the 
entropy pattern is significantly altered, resulting in a thermal wind balance 
consistent with the quenching of the winds 
\citep[e.g.][]{Showman06,Augustson12}.

Whether the magnetic effects are always required to confine the 
meridional circulation remains unclear. Indeed, in non-magnetic simulations,
viscous and thermal diffusion would mediate the penetration of the zonal winds 
into the SSL \citep{Spiegel92}. 
Given the large diffusivities adopted in global dynamo models, the penetration 
would be likely much more effective than realistic. In the context of 
modelling solar-type stars, \cite{Brun17} developed several non-magnetic 
numerical models in which the diffusivities are several orders of 
magnitude smaller in the SSL than in the convective envelope. This yields 
zonal flows that do not spread into the stably-stratified core (see their 
Fig.~11), at least on timescales smaller than the thermal diffusion time of the 
SSL.

The surface field in our simulation is too dipolar and shows too little 
localised field concentration when compared with the Jupiter field model JRM09 
by \cite{Connerney18}. The magnetic spectrum at the bottom of the stable 
layer at $0.84\,R_J$ is roughly compatible with JRM09 when upward-continued as 
a potential field. However, the skin effect and to a large degree also the 
dynamo action just above the stable layer heavily modifies the field, making it 
less realistic.

The efficiency of the dynamo action above the stable layer depends on the 
magnetic Reynolds number $Rm = U_z d_\sigma \sigma 
\mu_0 $ that is based on the zonal flow amplitude $U_z$
and the electrical conductivity scale height $d_\sigma=|\partial \ln \sigma / 
\partial r|^{-1}$ \citep{Liu08,Cao17,Wicht19a}. 
Observations of the magnetic field variations suggest that this magnetic 
Reynolds number, which increases with depth, reaches a value around unity
at $0.95\,R_J$ \citep{Moore19}. Gravity measurements indicate that this is also 
about the depth where the zonal wind velocity decreases rapidly 
\citep{Kaspi18,Galanti20}.

Additional support for the upper boundary comes from the fact that the width 
of the dominant equatorial jet on Jupiter ($\approx 30^\circ$) is only 
reproduced in numerical models when $\routs =0.95\,R_J$ 
\citep{Gastine14,Heimpel16}.

Recent interior models by \cite{Debras19} suggest  
a stably-stratified layer starting around $0.1$~Mbar, which 
would correspond to a somewhat deeper radius around $\routs=0.93\,R_J$.
However, the observational constraints (gravity, He abundance) likely also allow 
for a shallower layer. 
Helium demixing, considered as the best candidate to promote stable 
stratification in Jupiter, is expected where 
hydrogen becomes metallic and thus likely significantly deeper around 
$0.9\,R_J$. The possible physical origin of a stable layer that would 
start around $0.95\,R_J$ remains unclear.

The simulation presented here is the first to demonstrate that multiple zonal 
jets and Jupiter-like dynamo action can be consolidated in a global simulation. 
The necessary ingredient is a stably-stratified layer that allows zonal jets to 
develop in the outer envelope without contributing to the dynamo action 
in the deeper metallic region. 

While the simulation presented here is an important step towards more 
Jupiter-like models, there is certainly room for improvements. The simulation 
was performed at an Ekman number of $E=10^{-6}$
with considerable numerical costs. We speculate that an even smaller Ekman 
number, and possibly a larger Rayleigh number, is required to drive a stronger 
jet system that extends to yet higher latitudes. The magnetic field in our 
simulation could become more Jupiter-like for a stable layer that is 
thinner and lies closer to the surface than in our simulations. However, this 
would further increase the numerical costs. The magnetic field also lacks the 
characteristic banded structure that \cite{Gastine14a} attributed to zonal wind 
dynamo action. An increase of the conductivity, or rather the magnetic Reynolds 
number $Rm$, in the outer envelope in our simulation could help here. 
These open questions pave the way of future global Jovian dynamo models.

\section*{Acknowledgements}
We thank Dave Stevenson and an anonymous reviewer for their useful 
comments. Numerical computations have been carried out on the \texttt{S-CAPAD} 
platform at IPGP, on the \texttt{occigen} cluster at 
GENCI-CINES (Grant A0020410095) and on the \texttt{cobra} cluster in Garching.
All the figures have been generated using \texttt{matplotlib} \citep{Hunter07}
and \texttt{paraview} (\url{https://www.paraview.org}). The colormaps come from 
the \texttt{cmocean} package by \cite{cmocean}.

\appendix


\section{Approximations for the background state}
\label{sec:ref_coeffs}

To ensure that our numerical models could be computed again by other groups, we 
approximate the gravity profile $\gbar$, the expansion coefficient $\abar$ and 
the Gr\"uneisen parameter $\tilde{\Gamma}$ by simple interpolations. As shown 
in Fig.~\ref{fig:radprofs}, this yields a good agreement with the profile from 
\cite{French12}.

The dimensionless gravity profile is approximated by the following fourth-order 
polynomial of the dimensionless radius $r$

\begin{equation}
 \tilde{g}(r) = \sum_{i=0}^4 a^g_i r^i,
\end{equation}
where $a^g\simeq[0,2.435,0.162,-2.008, 0.665]$. The expansion coefficient is 
expressed by the following ninth-order polynomial

\begin{equation}
 \ln \abar(r)=\sum_{i=0}^9 a^\alpha_i r^i,
\end{equation}
where $a^\alpha\simeq 
[1.589,-1.228,4.532,-0.084,35.011,70.455,$ $-27.158,56.861
,64.873,-19.790 ]$.

The interior model of \cite{French12} suggests rapid variations of the 
Gr\"uneisen parameter in the outer 15\% of Jupiter (see 
Fig.~\ref{fig:profs}\textit{e}). It drops 
from a value of an almost constant value of $0.7-0.8$ in the metallic core down 
to $0.2$ around $0.9~R_J$. To account for this variation, we approximate the 
radial dependence of the dimensionless Gr\"uneisen parameter by a $\tanh$ 
function
\begin{equation}
  \tilde{\Gamma}(r)\simeq 1.313 -0.392\,\tanh\left[39.2(r-1.122)\right].
\end{equation}

\bibliographystyle{cas-model2-names}

\end{document}